\documentclass{SCAEOL}
\numberwithin{equation}{section}

\usepackage{tikz}
\usetikzlibrary{shapes,arrows}
\usetikzlibrary{arrows,shapes,chains}
\usepackage{bm}
\usepackage{graphicx} 
\usepackage{subfigure}
\usepackage{epstopdf}
\usepackage{tabularx}
\usepackage{multirow}
\usepackage{threeparttable}

\begin{document}

\Year{2020} %
\Month{Feburary}
\Vol{56} %
\No{1} %
\BeginPage{1} %
\EndPage{XX} %

\title{The impact of multilateral imported cases of COVID-19 on the epidemic control in China}{}


\author[1,4]{JIA Jiwei}{}
\author[2]{LIU Siyu}{Corresponding author}
\author[1]{DING Jian}{}
\author[1]{LIAO Guidong}{}
\author[3,4,5]{\\ZHANG Lihua}{}
\author[1,4]{ZHANG Ran}{}

\address[{\rm1}]{School of Mathematics, Jilin University, Changchun {\rm 130012};}
\address[{\rm2}]{School of Public Health, Jilin University, Changchun {\rm 130021};}
\address[{\rm3}]{Institute of AI \& Robotics, Fudan University, Shanghai, {\rm 200433};}
\address[{\rm4}]{Interdisciplinary Center of Jilin Province for Applied Mathematics, Changchun {\rm 130012};}
\address[{\rm5}]{Engineering Research Center of AI \& Unmanned Vehicle Systems of Jilin Province, Changchun, {\rm 130015};}

\Emails{jiajiwei@jlu.edu.cn (Jia J.), liusiyu@jlu.edu.cn (Liu S.), dingjian17@mails.jlu.edu.cn (Ding J.), liaogd18@mails.jlu.edu.cn (Liao G), lihuazhang@fudan.edu.cn (Zhang L.), zhangran@jlu.edu.cn (Zhang R.)}\maketitle


 {\begin{center}
\parbox{14.5cm}{\begin{abstract}
Nowadays, the epidemic of COVID-19 in China is under control. However, the epidemic are developing rapidly around the world. Due to the normal migration of population, China is facing high risk from imported cases. The potential specific medicine and vaccine is still in the process of clinical trials. Currently,  controlling the impact of imported cases is the key to prevent new outbreak of COVID-19 in China. In this paper, we propose two impulsive systems to describe the impact of multilateral imported cases of COVID-19. Based on the published data, we simulate and discussed the epidemic trends under different control strategies. We compare four different scenarios and show the corresponding medical burden. The results help to design appropriate control strategy for imported cases in practice.
\vspace{-3mm}
\end{abstract}}\end{center}}



\renewcommand{\baselinestretch}{1.2}

\baselineskip 11pt\parindent=10.8pt  \wuhao
\section{Introduction}

At the beginning of 2020, a horrible infectious disease spread rapidly throughout China. The strong ability of transmission and high fatality ratio have attracted attention worldwide. WHO named the infectious disease as Corona Virus Disease 2019 (COVID-19), it is the seventh member of coronavirus which can infect human \cite{WHOweb, Zhu2020}. Compared with Severe Acute Respiratory Syndrome (SARS) and Middle East Respiratory Syndrome (MERS), COVID-19 has a significant but volatile incubation period, a much stronger transmission ability and a relative low fatality ratio \cite{guanweijie}. Another key factor of COVID-19 is asymptomatic transmission,
it is reported \cite{nature} that the proportion of asymptomatic infected people could be up to $60\%$ and it brings high risk for new outbreak.
As of April 3, over 1 million confirmed cases were reported in more than 200 countries, areas or territories around the world \cite{WHOweb}.

The Chinese government has conducted a series strategies to prevent the spread of the disease, such as regional restrictions, home quarantine, isolating close contacts, extend vacation and postpone the resumption of work, these strategies are aiming at protecting the suspectable population \cite{nhc}. All provinces of China promptly launched Level-1 response to major public health emergencies and established designated hospitals, mobile cabin hospitals and fever outpatient departments to provide adequate medical support. Clinical studies showed that the disease is susceptible to all people, even worse, there is no specific medicine and vaccines. In order to ensure every patient to be treated, the Chinese government provide free medical care for domestic confirmed COVID-19 patients. In China, COVID-19 has been being under control by the strict measures since the end of February and the trend of the epidemic are becoming steady.

Although the number of cumulative infectious case nearly reached peak and the outbreak of the epidemic in China has been relieved greatly, the sporadic importations continue to affect the epidemic process. COVID-19 is now raging most region of the world. On March 17 \cite{WHOweb}, the total confirmed COVID-19 cases nearly reaches 100 thousands. Especially, Italy, Iran and Spain are all reported more than 10 thousands cases. Some of theses countries take similar measures with China to control the disease actively. Due to the international migration, the worldwide outbreak of COVID-19 still threaten China. How to keep up our costly achievement is the most important topic for the epidemic control. In next control stage, we will focus on the management and controlling of cases imported from foreign countries.

Extensive researches for COVID-19 with multiple points of view has been reported. These researches make people understand COVID-19 better. The study of clinical observation and management of COVID-19 patients will be the guideline of practical treatment \cite{chenhuijun,gaojianjun,guanweijie}. The development of specific medicine and vaccines is mainly depended on virological research \cite{panyang,wrapp,zoulirong}. Further more, most people want to know the trend and scale of the disease. Appropriate prediction will help people avoid more loss and make effective control strategy \cite{prem,shenmingwang,tianhuaiyu,zhangsheng}.

We incorporate impulsive terms into differential equations to simulate the process of migration from other countries. We discuss several different control strategies under different scenarios and estimate the accumulated medical resource for each procedure. Currently, most of provinces in China are resuming to work, we analyze the impact of imported cases with and without the domestic cases cleared under the different quarantine strategies. It shows that the imported cases would produce new outbreak of the epidemic if the domestic infected cases were not clear. After clearing that, the increasing trend due to the imported cases can be suppressed under strict immigration control strategy. The numerical results also show that limitation of international arrivals, the strategy carrying in China, is an effective way to reduce the amount of confirmed cases and also the medical resources needed.

The rest of the paper is organized as follows. We propose the impulsive models in Section 2 and present the numerical results in Section 3.
Finally, we present some concluding remarks in Section 4. Some supplementary materials are shown in Appendix.

\section{Model Formulation}
In this section, we formulate the epidemic model in an impulsive way to describe multilateral immigration. The epidemic of COVID-19 is under control in most provinces in China, some provinces have been preparing for resumption of work orderly. The complete recovery of production and normal daily life is coming soon. Based on our previous work in \cite{EJDE} and \cite{SCMcn}, we propose two models to describe the impact of imported cases with and without the original quarantine strategy applied. The models satisfy the following assumptions.

\begin{itemize}
\item[(1)] All coefficients involved in the models are positive constants, $Z_{+}$~represents the set of positive integer.
\item[(2)] Natural birth and death are not taken into account.
\item[(3)] Once the infected patient is cured, the immune efficacy will maintain for some time, i.e., second infection is not considered in the model.
\end{itemize}

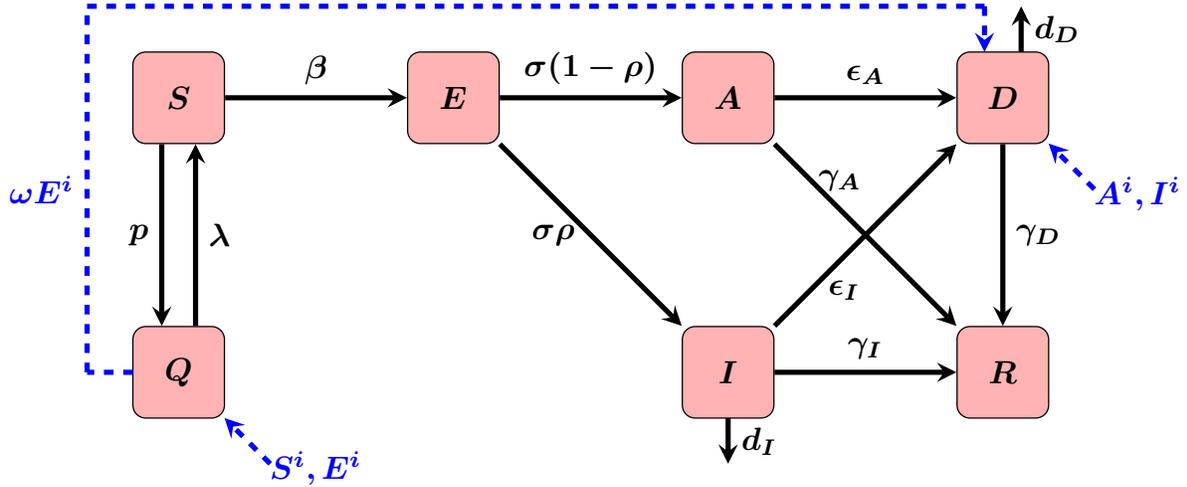
\begin{figure}[h!]
\centering
\tikzstyle{startstop} = [rectangle,rounded corners, minimum width=1cm,minimum height=1cm,text centered, draw=black,fill=red!30]
\tikzstyle{style_S} = [rectangle,rounded corners, minimum width=1cm,minimum height=1cm,text centered, draw=black,fill=red!30]
\tikzstyle{style_Q} = [rectangle,rounded corners, minimum width=1cm,minimum height=1cm,text centered, draw=black,fill=red!30]
\tikzstyle{style_E} = [rectangle,rounded corners, minimum width=1cm,minimum height=1cm,text centered, draw=black,fill=red!30]
\tikzstyle{style_A} = [rectangle,rounded corners, minimum width=1cm,minimum height=1cm,text centered, draw=black,fill=red!30]
\tikzstyle{style_I} = [rectangle,rounded corners, minimum width=1cm,minimum height=1cm,text centered, draw=black,fill=red!30]
\tikzstyle{style_D} = [rectangle,rounded corners, minimum width=1cm,minimum height=1cm,text centered, draw=black,fill=red!30]
\tikzstyle{style_R} = [rectangle,rounded corners, minimum width=1cm,minimum height=1cm,text centered, draw=black,fill=red!30]
\tikzstyle{arrow} = [ultra thick,->,>=stealth]
\tikzstyle{line} = [ultra thick,-,>=stealth]
\resizebox{\textwidth}{!}{
\begin{tikzpicture}[node distance=2cm]
\node(S)[style_S]{$\bm S$};
\node(Q)[style_Q,below of=S,yshift=-1.cm]{$\bm Q$};
\node(E)[style_E,right of =S,xshift=1.cm]{$\bm E$};
\node(A)[style_A,right of =E,xshift=1.cm]{$\bm A$};
\node(I)[style_I,below of=A,yshift=-1.cm]{$\bm I$};
\node(D)[style_D,right of =A,xshift=1.cm]{$\bm D$};
\node(R)[style_R,below of=D,yshift=-1.cm]{$\bm R$};
\draw [arrow] (S.-110) -- (Q.110) node[midway,left] {$\bm p$};
\draw [arrow] (Q.70) -- (S.-70) node[midway, right] {$\bm\lambda$};
\draw [arrow] (S) -- (E) node[midway,above] {$\bm\beta$};
\draw [arrow] (E) -- (A)node[midway,above] {$\bm{\sigma(1-\rho)}$};
\draw [arrow] (E) -- (I)node[midway,left,xshift=-0.3mm] {$\bm{\sigma\rho}$};
\draw [arrow] (A) -- (D)node[midway,above] {$\bm{\epsilon_{A}}$};
\draw [arrow] (A) -- (R)node[midway,left,yshift=0.6cm,xshift=0.1cm] {$\bm{\gamma_{A}}$};
\draw [arrow] (I) -- (D)node[midway,right,yshift=-0.6cm,xshift=-0.55cm] {$\bm{\epsilon_{I}}$};
\draw [arrow] (I) -- (R)node[midway,above] {$\bm{\gamma_{I}}$};
\draw [arrow] (D) -- (R)node[midway,right] {$\bm{\gamma_{D}}$};
\draw [arrow] (6,-3.5) -- (6,-4)node[midway,right] {$\bm{d_{I}}$};
\draw [arrow] (9.2,0.5) -- (9.2,1)node[midway,right] {$\bm{d_{D}}$};
\draw [blue,dashed,arrow] (1,-4) -- (0.5,-3.5)node[midway,right,yshift=-0.3cm,xshift=0.1cm] {$\bm{S^{i}, E^{i}}$};
\draw [blue,dashed,arrow] (10,-1) -- (9.5,-0.5)node[midway,right,yshift=-0.3cm,xshift=0.1cm] {$\bm{A^{i}, I^{i}}$};
\draw [blue,dashed,line] (-0.5,-3) -- (-1,-3);
\draw [blue,dashed,line] (-1,-3) -- (-1,1)node[midway,left] {$\bm{\omega E^{i}}$};
\draw [blue,dashed,line] (-1,1) -- (8.8,1);
\draw [blue,dashed,arrow] (8.8,1) -- (8.8,0.5);
\end{tikzpicture}}
\caption{Flow diagram of Model~(\ref{model1})}\label{fig_illustration1}
\end{figure}

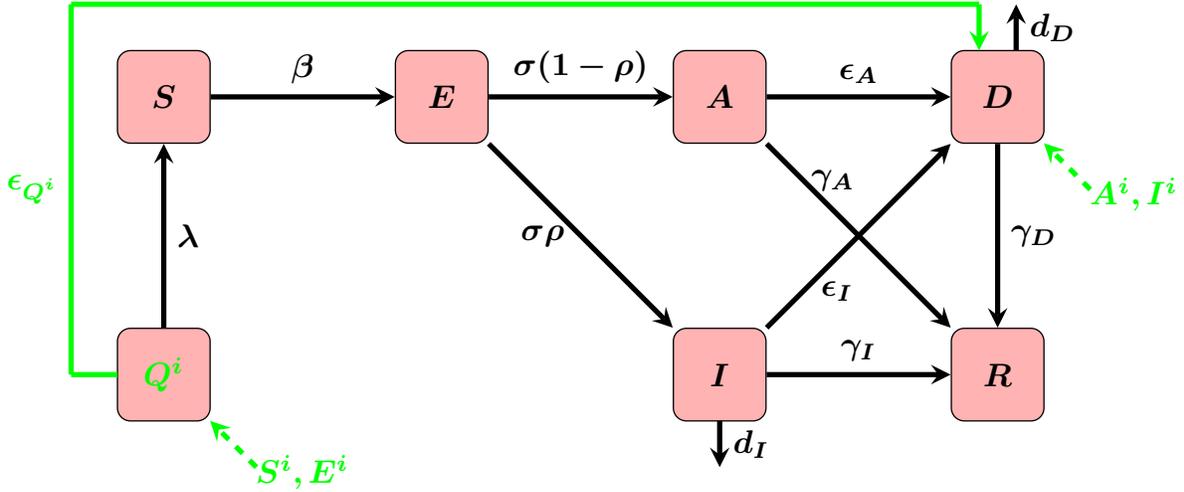
\begin{figure}[h!]
\centering
\tikzstyle{startstop} = [rectangle,rounded corners, minimum width=1cm,minimum height=1cm,text centered, draw=black,fill=red!30]
\tikzstyle{style_S} = [rectangle,rounded corners, minimum width=1cm,minimum height=1cm,text centered, draw=black,fill=red!30]
\tikzstyle{style_Q} = [rectangle,rounded corners, minimum width=1cm,minimum height=1cm,text centered, draw=black,fill=red!30]
\tikzstyle{style_E} = [rectangle,rounded corners, minimum width=1cm,minimum height=1cm,text centered, draw=black,fill=red!30]
\tikzstyle{style_A} = [rectangle,rounded corners, minimum width=1cm,minimum height=1cm,text centered, draw=black,fill=red!30]
\tikzstyle{style_I} = [rectangle,rounded corners, minimum width=1cm,minimum height=1cm,text centered, draw=black,fill=red!30]
\tikzstyle{style_D} = [rectangle,rounded corners, minimum width=1cm,minimum height=1cm,text centered, draw=black,fill=red!30]
\tikzstyle{style_R} = [rectangle,rounded corners, minimum width=1cm,minimum height=1cm,text centered, draw=black,fill=red!30]
\tikzstyle{arrow} = [ultra thick,->,>=stealth]
\tikzstyle{line} = [ultra thick,-,>=stealth]
\resizebox{\textwidth}{!}{
\begin{tikzpicture}[node distance=2cm]
\node(S)[style_S]{$\bm S$};
\node(Q)[style_Q,below of=S,yshift=-1.cm]{$\bm{{\color{green}Q^{i}}}$};
\node(E)[style_E,right of =S,xshift=1.cm]{$\bm E$};
\node(A)[style_A,right of =E,xshift=1.cm]{$\bm A$};
\node(I)[style_I,below of=A,yshift=-1.cm]{$\bm I$};
\node(D)[style_D,right of =A,xshift=1.cm]{$\bm D$};
\node(R)[style_R,below of=D,yshift=-1.cm]{$\bm R$};
\draw [arrow] (Q.north) -- (S.south) node[midway, right] {$\bm\lambda$};
\draw [arrow] (S) -- (E) node[midway,above] {$\bm\beta$};
\draw [arrow] (E) -- (A)node[midway,above] {$\bm{\sigma(1-\rho)}$};
\draw [arrow] (E) -- (I)node[midway,left,xshift=-0.3mm] {$\bm{\sigma\rho}$};
\draw [arrow] (A) -- (D)node[midway,above] {$\bm{\epsilon_{A}}$};
\draw [arrow] (A) -- (R)node[midway,left,yshift=0.6cm,xshift=0.1cm] {$\bm{\gamma_{A}}$};
\draw [arrow] (I) -- (D)node[midway,right,yshift=-0.6cm,xshift=-0.55cm] {$\bm{\epsilon_{I}}$};
\draw [arrow] (I) -- (R)node[midway,above] {$\bm{\gamma_{I}}$};
\draw [arrow] (D) -- (R)node[midway,right] {$\bm{\gamma_{D}}$};
\draw [arrow] (6,-3.5) -- (6,-4)node[midway,right] {$\bm{d_{I}}$};
\draw [arrow] (9.2,0.5) -- (9.2,1)node[midway,right] {$\bm{d_{D}}$};
\draw [green,line] (-0.5,-3) -- (-1.0,-3);
\draw [green,line] (-1,-3) -- (-1,1)node[midway,left] {$\bm{\epsilon_{Q^{i}}}$};
\draw [green,line] (-1,1) -- (8.8,1);
\draw [green,arrow] (8.8,1) -- (8.8,0.5);
\draw [green,dashed,arrow] (1,-4) -- (0.5,-3.5)node[midway,right,yshift=-0.3cm,xshift=0.1cm] {$\bm{S^{i}, E^{i}}$};
\draw [green,dashed,arrow] (10,-1) -- (9.5,-0.5)node[midway,right,yshift=-0.3cm,xshift=0.1cm] {$\bm{A^{i}, I^{i}}$};
\end{tikzpicture}}
\caption{Flow diagram of Model~(\ref{model2})}\label{fig_illustration2}
\end{figure}

The flow diagrams (Figure~\ref{fig_illustration1} and \ref{fig_illustration2}) present the control strategy. Solid and dashed lines represent the continuous and the impulsive importing procedures, respectively. The corresponding dynamical systems are shown in Model (\ref{model1}) and (\ref{model2}). For Model (\ref{model1}), we adopt the Immigration Strategy III in~\cite{SCMcn}, which is carrying out in many ports of entry, such as Beijing, Shanghai and Guangzhou etc.

\begin{equation}\label{model1}
\left\{
    \begin{array}{l}
        \left.
        \begin{array}{ccl}
        \dfrac{{\rm{d}} S}{{\rm{d}} t}&=&-\beta S(I+\theta A) - pS + \lambda Q\\[3mm]
        \dfrac{{\rm{d}}   Q}{{\rm{d}} t}&=& pS-\lambda Q\\[3mm]
        \dfrac{{\rm{d}} E}{{\rm{d}} t}&=& \beta S(I + \theta A) - \sigma E\\[3mm]
        \dfrac{{\rm{d}}  A}{{\rm{d}} t}&=&\sigma(1-\rho)E - \epsilon_{A}A-\gamma_{A}A\\[3mm]
        \dfrac{{\rm{d}}  I}{{\rm{d}} t}&=&\sigma\rho E - \gamma_{I}I - d_{I}I - \epsilon_{I}I\\[3mm]
        \dfrac{{\rm{d}} D}{{\rm{d}} t}&=&\epsilon_{A}A + \epsilon_{I}I - d_{D}D - \gamma_{D}D\\[3mm]
        \dfrac{{\rm{d}} R}{{\rm{d}} t}&=&\gamma_{A}A + \gamma_{I}I + \gamma_{D}D\\[4mm]
        \end{array},
        \right\}t\neq nT,\\[3mm]
        \left.\begin{array}{ccl}
        Q(t^{+})&=&Q(t)+ S^{i} + E^{i} - \omega E^{i},\\[3mm]
        D(t^{+})&=&D(t)+ A^{i}+I^{i} + \omega E^{i}.
        \end{array}\right\}  t=nT,\ n\in Z_{+}.
    \end{array}
\right.
\end{equation}
For Model (\ref{model2}), we propose a new strategy, under which all the people inside China resume to work completely and strict isolation for the imported population from other countries. The detail process is shown in Figure \ref{fig_illustration2}.

\begin{equation}\label{model2}
\left\{
    \begin{array}{l}
        \left.
        \begin{array}{ccl}
        \dfrac{{\rm{d}} S}{{\rm{d}} t}&=&-\beta S(I+\theta A) + \lambda Q^{i}\\[3mm]
        \dfrac{{\rm{d}}   Q^{i}}{{\rm{d}} t}&=&-\lambda Q^{i}- \epsilon_{Q^{i}}Q^{i}\\[3mm]
        \dfrac{{\rm{d}} E}{{\rm{d}} t}&=& \beta S(I + \theta A) - \sigma E\\[3mm]
        \dfrac{{\rm{d}}  A}{{\rm{d}} t}&=&\sigma(1-\rho)E - \epsilon_{A}A-\gamma_{A}A\\[3mm]
        \dfrac{{\rm{d}}  I}{{\rm{d}} t}&=&\sigma\rho E - \gamma_{I}I - d_{I}I - \epsilon_{I}I\\[3mm]
        \dfrac{{\rm{d}} D}{{\rm{d}} t}&=&\epsilon_{A}A + \epsilon_{I}I - d_{D}D - \gamma_{D}D\\[3mm]
        \dfrac{{\rm{d}} R}{{\rm{d}} t}&=&\gamma_{A}A + \gamma_{I}I + \gamma_{D}D\\[4mm]
        \end{array},
        \right\}t\neq nT,\\[3mm]
        \left.\begin{array}{ccl}
        Q^{i}(t^{+})&=&Q^{i}(t)+ S^{i} + E^{i},\\[3mm]
        D(t^{+})&=&D(t)+ A^{i}+I^{i}.
        \end{array}\right\}  t=nT,\ n\in Z_{+}.
    \end{array}
\right.
\end{equation}
Where $S(t), Q(t) (Q^{i}(t)), E(t), A(t), I(t), D(t)$ and $R(t)$ denotes the susceptible, quarantined, exposed, asymptomatically infected, symptomatically infected, diagnosed and recovered population at time $t$, respectively. Compartment $E(t)$ represents low-level virus carriers, which are considered to be no infectiousness. We assume that the individuals in compartment $D(t)$ are being treated and isolated. In Model (\ref{model1}), the home quarantine strategy keeps carrying on, we use parameter $p$ and $\lambda$ to represent the quarantined rate and release rate. Compartment $Q(t)$ in Model~(\ref{model1}) contains not only the imported exposed population $E^{i}$,
but also the domestic quarantined individuals who are carrying no virus, so we use an impulsive procedure to describe the transition of $E^{i}$ to $D(t)$ for Model~(\ref{model1}). $\omega$ denotes the proportion of becoming diagnosed from the imported exposed population $E^{i}$ in Model~(\ref{model1}) during the isolation period. Model~(\ref{model2}) describes the scenario after the resumption of work, all of the domestic population in China return back to their normal life, then the compartment $Q^{i}(t)$ includes the isolated imported population only.  We use a continuous transition procedure and denote by parameter $\epsilon_{Q^{i}}$  the diagnostic rate of imported exposed population in the model. Both $Q(t)$ and $Q^{i}(t)$ have no contact with the infected individuals, that is, the population in $Q(t)$ and $Q^{i}(t)$ will not cause new infection. However, there are some exposed individuals in them due to the overseas imported cases, which we cannot detect them immediately after arrival. $S^{i}, E^{i}, A^{i}$ and $I^{i}$ are constants which denote the population of corresponding compartments from the imported population, the period is denoted as $T$.

Model~(\ref{model1}) and (\ref{model2}) share $11$ parameters, we refer the definition of these parameters in~\cite{EJDE}. We use parameter $\beta$ to denote the contact rate and $\theta\in (0,1)$ to denote the ratio between the infection rates of the individuals with and without symptoms. $\sigma$ denotes the transition rate of exposed to infected class. After someone infected, the proportion of becoming symptomatic is denoted by $\rho$ and asymptomatic by $1-\rho$. Diagnostic rate of asymptomatic and symptomatic infectious are respectively denoted by $\epsilon_{A}$ and $\epsilon_{I}$ and the mean recovery period of class $A, I, D$ are denoted by $1/\gamma_{A}, 1/\gamma_{I}$ and $1/\gamma_{D}$, respectively. The parameters $d_{I}$ and $d_{D}$ represent the disease-induced death rate.

For the two models established in this paper, overseas cases import into China periodically. The continuous input will make $D(t)>0$ at any time. We define this state as fake diseases-free status. With meticulous isolation and control, it may not cause new local infected. However, it requires strict immigration policy and must pay more attention to avoid nosocomial infection. The medical burden of COVID-19 will be lasting long. Once infected individuals come into our normal social life, it might cause a more serious new outbreak which we show it in Section 3. The epidemic vanishes only if there are no imported cases. The world is a community of common destiny for all mankind, therefore, China must prepare for long-term actions on the prevention of COVID-19 imported cases.

\section{Numerical Simulations}
According to the severity of the epidemic around the world and the closeness of personal exchange with China,
we take Italy, Iran, Spain, Germany and France - the five countries with the most sever COVID-19 outbreaks, South Korea and Japan -
the two neighboring countries, into account (\cite{WHOweb}, the data updated to March 17, 2020). We select 16 provinces in China, which have the relatively large numbers of imported population from these countries. We simulate the impact of the imported population from these seven countries. The control strategy is changing with the process of epidemic, we forecast the trends of COVID-19 under different scenarios.

\subsection{Estimation of the Imported Population}
The sources of imported population is relatively fixed. Based on the international flight schedule published by Civil Aviation Administration of China (CAAC)~\cite{minhang} and the ship information on China Ship Ticketing Website~\cite{chuanpiao}, we obtained the departure, destination, the weekly flights and ships information. Combining with the epidemic data of COVID-19 published by WHO~\cite{WHOweb}, we estimate the population imported from different airports and ports of the selected seven countries, also the arrivals of different compartments. Table~\ref{table_import1} and \ref{table_import2} in Appendix show the detailed imported population and the estimated proportion of different compartments from different airports and ports are in Table \ref{table_proportion}. Airport codes are defined by International Air Transport Association~(IATA)~\cite{IATAweb}, port codes are according to the China National Standard GB/T 7407-2015~\cite{ports}.
From Table~\ref{table_import1} and \ref{table_import2}, we can obtain the periodical impulsive population $S^{i}, E^{i}, A^{i}, I^{i}$ for simulation, we assume the impulsive period $T = 3$ days.

\renewcommand\arraystretch{1.3}
\begin{table}[htp]
\centering
\resizebox{0.8\textwidth}{!}{
\begin{threeparttable}
\caption{Proportion of $E, A$ and $I$ in imported population}\label{table_proportion}
\begin{tabular}{c|c|c|c|c|c|c|c|c|c}
\hline
\hline
\textbf{Country}&\textbf{A/P$^{1}$}&	$\bm{E^{i}}$	&	$\bm{A^{i}}$	&	$\bm{I^{i}}$	&\textbf{Country}&\textbf{A/P$^{1}$}&	$\bm{E^{i}}$	&	$\bm{A^{i}}$	&	$\bm{I^{i}}$	\\
\hline
\multirow{5}{*}{Korea}	&	ICN	&	0.90\%	&	0.03\%	&	0.08\%	&	Iran	&	IKA	&	2.94\%	&	0.09\%	&	0.16\%	\\
\cline{6-10}
	&	GMP	&	1.80\%	&	0.06\%	&	0.12\%	&	France	&	CDG	&	1.80\%	&	0.06\%	&	0.12\%	\\
\cline{6-10}
	&	CJU	&	0.90\%	&	0.03\%	&	0.08\%	&	\multirow{2}{*}{Germany}	&	FRA	&	2.20\%	&	0.08\%	&	0.14\%	\\
	&	PUS	&	1.80\%	&	0.06\%	&	0.12\%	&		&	MUC	&	2.94\%	&	0.09\%	&	0.16\%	\\
\cline{6-10}
	&	TAE	&	2.94\%	&	0.09\%	&	0.16\%	&	\multirow{2}{*}{Spain}	&	BCN	&	2.20\%	&	0.08\%	&	0.14\%	\\
\cline{1-5}
\multirow{10}{*}{Japan}	&	KIX	&	2.20\%	&	0.08\%	&	0.14\%	&		&	MAD	&	2.94\%	&	0.09\%	&	0.16\%	\\
\cline{6-10}
	&	HND	&	2.20\%	&	0.08\%	&	0.14\%	&	\multirow{2}{*}{Italy}	&	FCO	&	1.80\%	&	0.06\%	&	0.12\%	\\
	&	NGO	&	2.20\%	&	0.08\%	&	0.14\%	&		&	LIN	&	2.94\%	&	0.09\%	&	0.16\%	\\
\cline{6-10}
	&	CTS	&	2.20\%	&	0.08\%	&	0.14\%	&	\multirow{3}{*}{Korea}	&	SKINC	&	0.90\%	&	0.03\%	&	0.08\%	\\
	&	NRT	&	2.20\%	&	0.08\%	&	0.14\%	&		&	KRPTK	&	1.80\%	&	0.06\%	&	0.12\%	\\
	&	FUK	&	0.90\%	&	0.03\%	&	0.08\%	&		&	SKKUN	&	0.90\%	&	0.03\%	&	0.08\%	\\
\cline{6-10}
	&	HIJ	&	0.90\%	&	0.03\%	&	0.08\%	&	\multirow{3}{*}{Japan}	&	JPOSK	&	2.20\%	&	0.08\%	&	0.14\%	\\
	&	KMQ	&	0.90\%	&	0.03\%	&	0.08\%	&		&	JPKOB	&	1.80\%	&	0.06\%	&	0.12\%	\\
	&	SDJ	&	0.90\%	&	0.03\%	&	0.08\%	&		&	JPSHI	&	0.90\%	&	0.03\%	&	0.08\%	\\
\cline{6-10}
	&	OKA	&	0.90\%	&	0.03\%	&	0.08\%	&		&		&		&		&		\\

\hline
\hline
\end{tabular}
\begin{tablenotes}
\footnotesize
\item[1]A/P: airports or ports.
\end{tablenotes}
\end{threeparttable}}
\end{table}

\subsection{Accumulated Medical Resource}
We discussed the long-term problem of medical burden which China has to face in Section 2.
In this section, we introduce an index to estimate the medical burden, which reflects the prevalence of COVID-19 and helps to make design for the allocation of health care resources.
Following the procedure in \cite{EJDE}, we integrate $D(t)$ in
$[t_{0}, t_{1}]$ to represent $AMR$ in this period.
\begin{equation}\label{AMR}
AMR = r\int_{t_{0}}^{{t_{1}}}D(t){\rm{d}}t,
\end{equation}
where $r$ represents the average medical resource a patient needs daily.

\subsection{Numerical Results}
In this section, we simulate the impact of the imported population under the following 4 scenarios:
\begin{itemize}
\item[A.]Home quarantine strategy keeps carrying and the imported population is in normal size.
\item[B.]Resume to work without clearing the domestic infected cases and the imported population is in normal size.
\item[C.]Resume to work after clearing the domestic infected cases and the imported population is in normal size
\item[D.]Resume to work after clearing the domestic infected cases and the arrival population is limited.
\end{itemize}

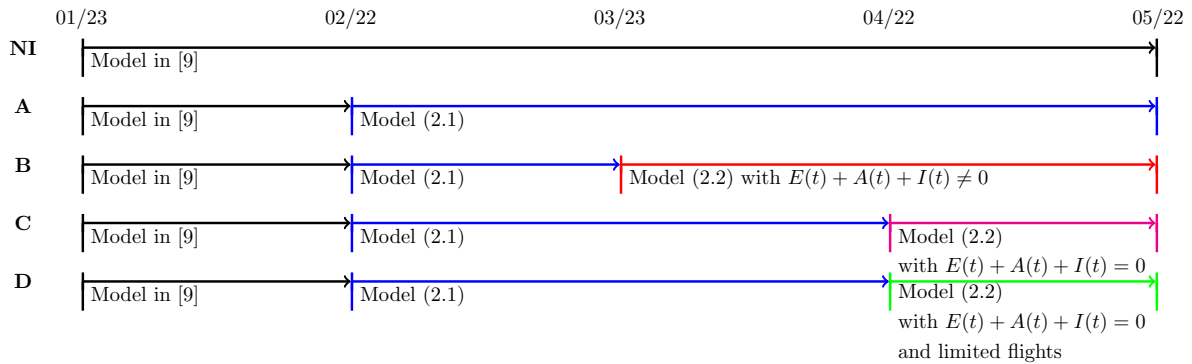
\begin{figure}[h!]
\centering
\tikzstyle{S123} = [rectangle,rounded corners, minimum width=1cm,minimum height=1cm]
\tikzstyle{arrow} = [ultra thick,->,>=stealth]
\tikzstyle{line} = [ultra thick,-,>=stealth]
\resizebox{1\textwidth}{!}{
\begin{tikzpicture}[xscale = 20]
\node(N)[S123]{\textbf{NI}};
\node(I)[S123,below of = N]{\textbf{A}};
\node(II)[S123, below of = I]{\textbf{B}};
\node(III)[S123, below of = II]{\textbf{C}};
\node(III)[S123, below of = III]{\textbf{D}};
\draw[|->|][draw=black, very thick] (0.05,0) -- (0.97,0);
\draw[|->][draw=black, very thick] (0.05,-1) -- (0.28,-1);
\draw[|->|][draw=blue, very thick] (0.28,-1) -- (0.97,-1);
\draw[|->][draw=black, very thick] (0.05,-2) -- (0.28,-2);
\draw[|->][draw=blue, very thick] (0.28,-2) -- (0.51,-2);
\draw[|->|][draw=red, very thick] (0.51,-2) -- (0.97,-2);
\draw[|->][draw=black, very thick] (0.05,-3) -- (0.28,-3);
\draw[|->][draw=blue, very thick] (0.28,-3) -- (0.74,-3);
\draw[|->|][draw=magenta, very thick] (0.74,-3) -- (0.97,-3);
\draw[|->][draw=black, very thick] (0.05,-4) -- (0.28,-4);
\draw[|->][draw=blue, very thick] (0.28,-4) -- (0.74,-4);
\draw[|->|][draw=green, very thick] (0.74,-4) -- (0.97,-4);
\draw[-][draw=black,very thick] (0.051,0) -- (0.051,-0.5) node[midway,right]{Model in \cite{EJDE}};
\draw[-][draw=black,very thick] (0.969,0) -- (0.969,-0.5);
\draw[-][draw=black,very thick] (0.051,-1) -- (0.051,-1.5) node[midway,right]{Model in \cite{EJDE}};
\draw[-][draw=black,very thick] (0.051,-2) -- (0.051,-2.5) node[midway,right]{Model in \cite{EJDE}};
\draw[-][draw=black,very thick] (0.051,-3) -- (0.051,-3.5) node[midway,right]{Model in \cite{EJDE}};
\draw[-][draw=black,very thick] (0.051,-4) -- (0.051,-4.5) node[midway,right]{Model in \cite{EJDE}};
\draw[-][draw=blue,very thick] (0.281,-1) -- (0.281,-1.5) node[midway,right]{Model~(\ref{model1})};
\draw[-][draw=blue,very thick] (0.969,-1) -- (0.969,-1.5);
\draw[-][draw=blue,very thick] (0.281,-2) -- (0.281,-2.5) node[midway,right]{Model~(\ref{model1})};
\draw[-][draw=red,very thick] (0.511,-2) -- (0.511,-2.5) node[midway,right]{Model~(\ref{model2}) with $E(t)+A(t)+I(t)\neq0$};
\draw[-][draw=red,very thick] (0.969,-2) -- (0.969,-2.5);
\draw[-][draw=blue,very thick] (0.281,-3) -- (0.281,-3.5) node[midway,right]{Model~(\ref{model1})};
\draw[-][draw=blue,very thick] (0.281,-4) -- (0.281,-4.5) node[midway,right]{Model~(\ref{model1})};
\draw[-][draw=magenta,very thick] (0.741,-3) -- (0.741,-3.5)node[align=left, right]{Model~(\ref{model2})\\ with $E(t)+A(t)+I(t)=0$};
\draw[-][draw=green,very thick] (0.741,-4) -- (0.741,-4.5)node[align=left, right,yshift = -0.2cm]{Model~(\ref{model2})\\ with $E(t)+A(t)+I(t)=0$\\ and limited flights};
\draw[-][draw=magenta,very thick] (0.969,-3) -- (0.969,-3.5);
\draw[-][draw=green,very thick] (0.969,-4) -- (0.969,-4.5);
\node[align=center]at(0.05,0.5){01/23};
\node[align=center]at(0.28,0.5){02/22};
\node[align=center]at(0.51,0.5){03/23};
\node[align=center]at(0.74,0.5){04/22};
\node[align=center]at(0.97,0.5){05/22};
\end{tikzpicture}}
\caption{Illustration for simulation procedure of Scenario A, B, C and D. (NI: No imported)}\label{fig_flow123}
\end{figure}

Figure~\ref{fig_flow123} illustrates the simulation procedure for above scenarios. We divide the simulation period, from Jan 23 to May 22, into four parts,
with $30$ days for each part. Based on current epidemic situation in 
\begin{figure}[htp!]
\centering
\subfigure{
\begin{minipage}[t]{0.25\textwidth}
\centering
\includegraphics[width=3.6cm]{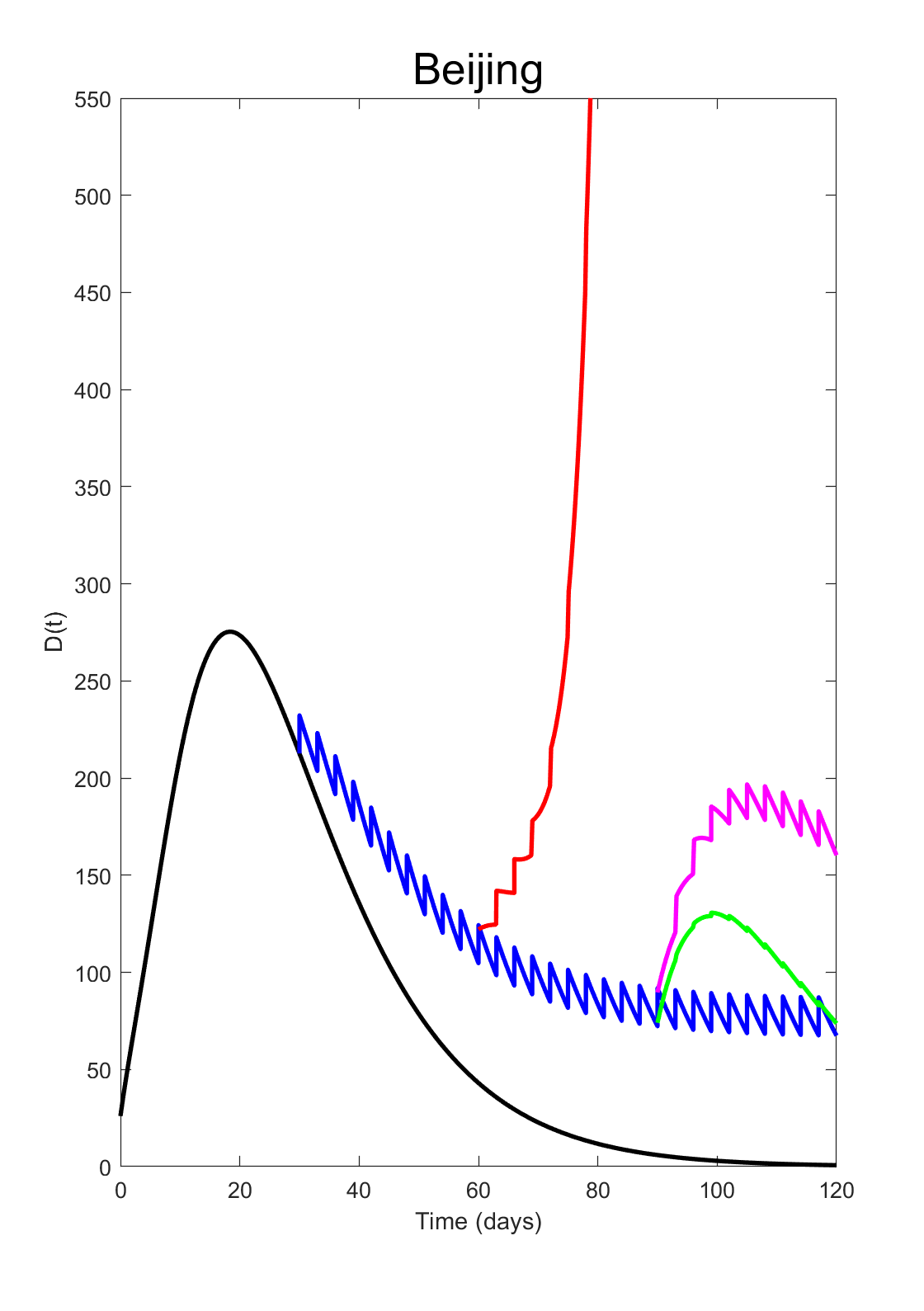}
\end{minipage}%
}%
\subfigure{
\begin{minipage}[t]{0.25\textwidth}
\centering
\includegraphics[width=3.6cm]{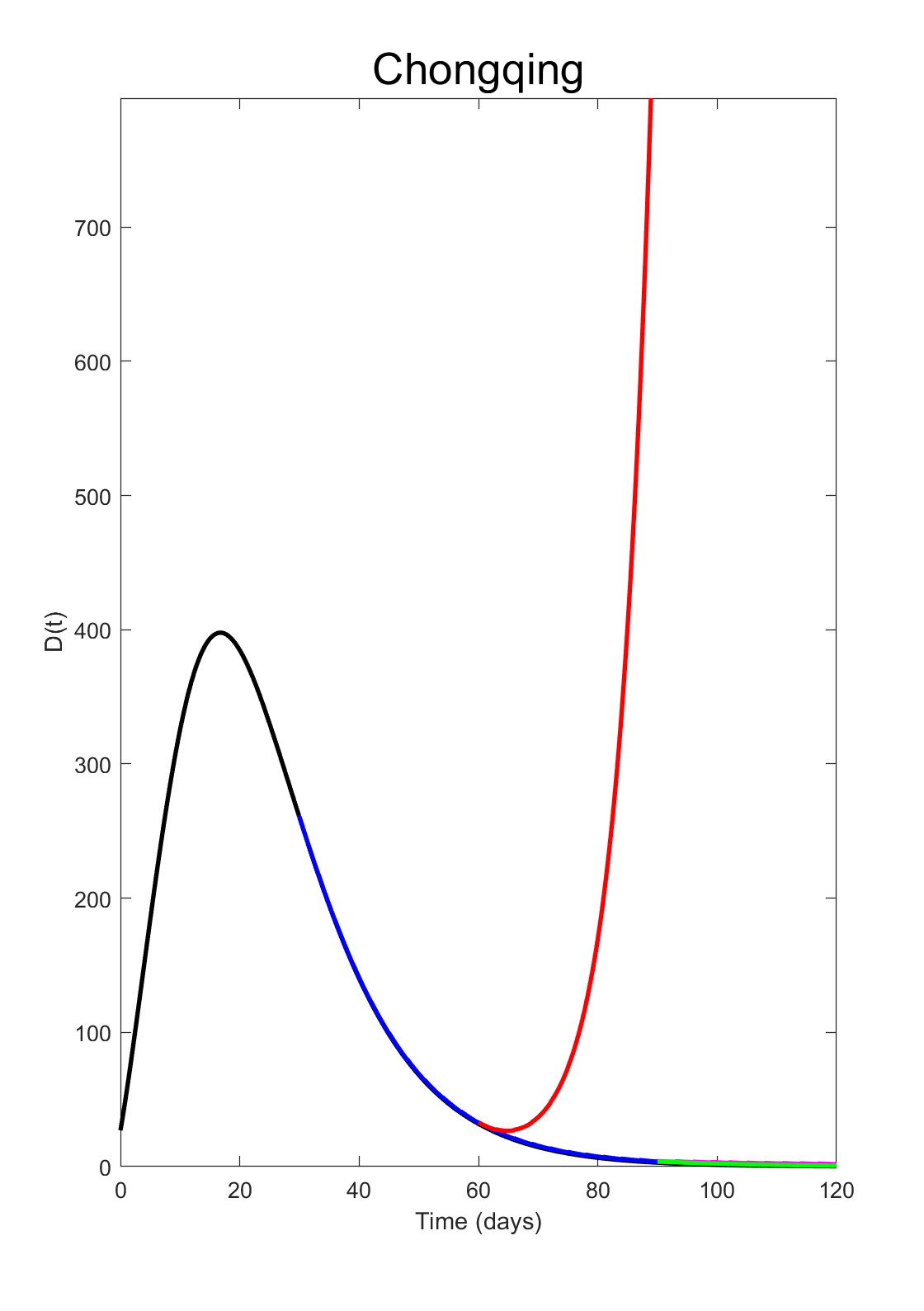}
\end{minipage}%
}%
\subfigure{
\begin{minipage}[t]{0.25\textwidth}
\centering
\includegraphics[width=3.6cm]{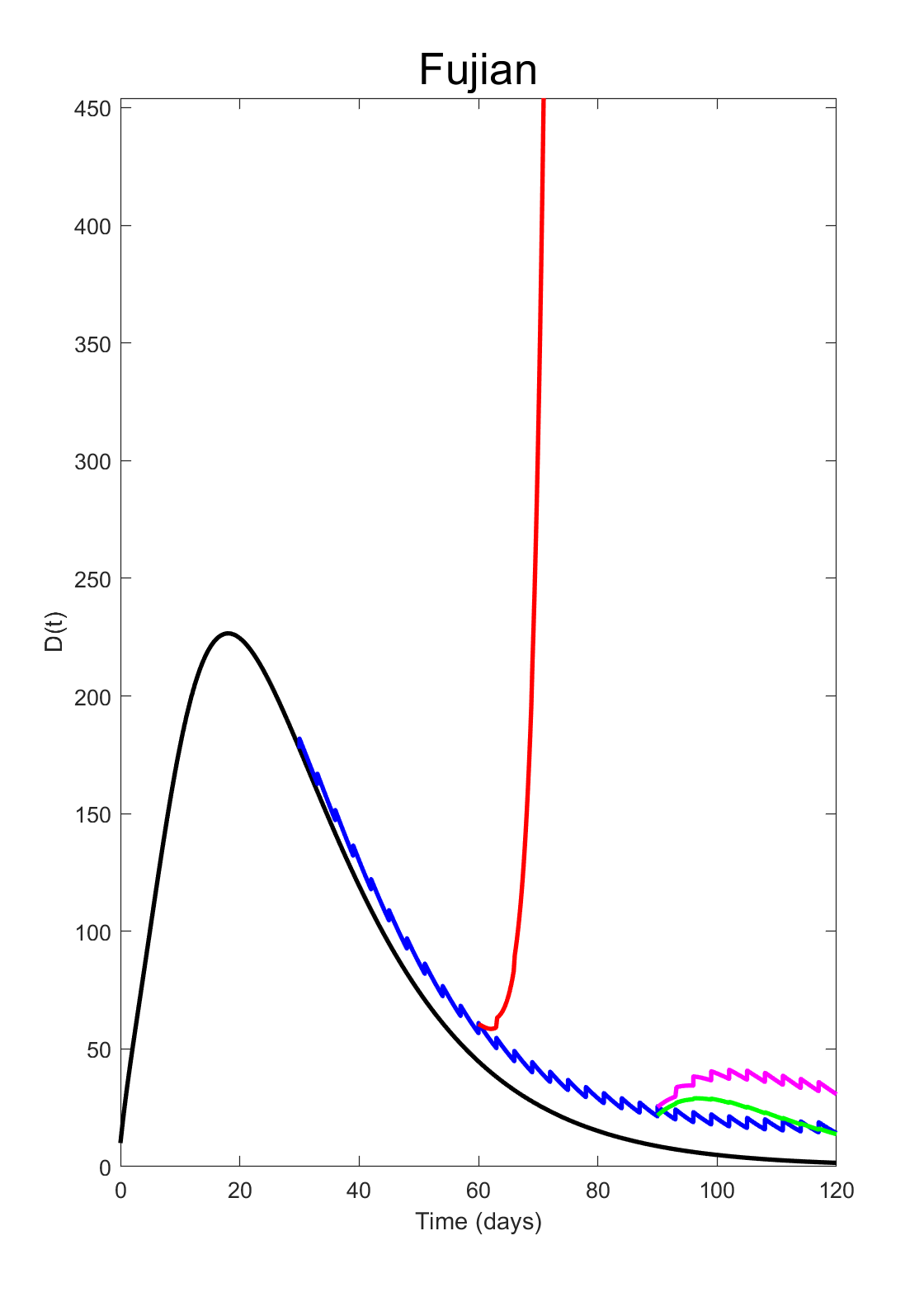}
\end{minipage}%
}%
\subfigure{
\begin{minipage}[t]{0.25\textwidth}
\centering
\includegraphics[width=3.6cm]{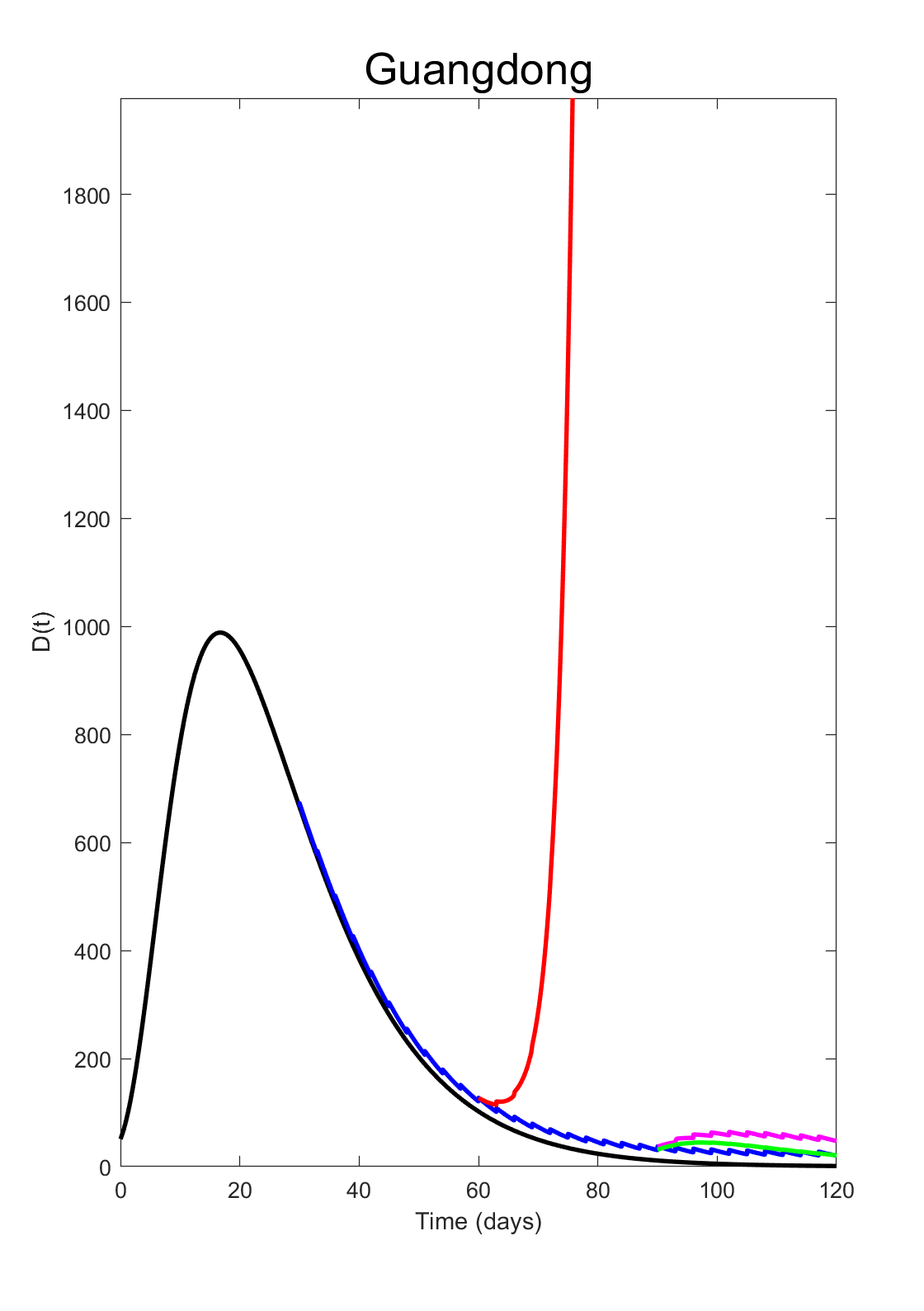}
\end{minipage}%
}%
\vskip -10pt

\subfigure{
\begin{minipage}[t]{0.25\textwidth}
\centering
\includegraphics[width=3.6cm]{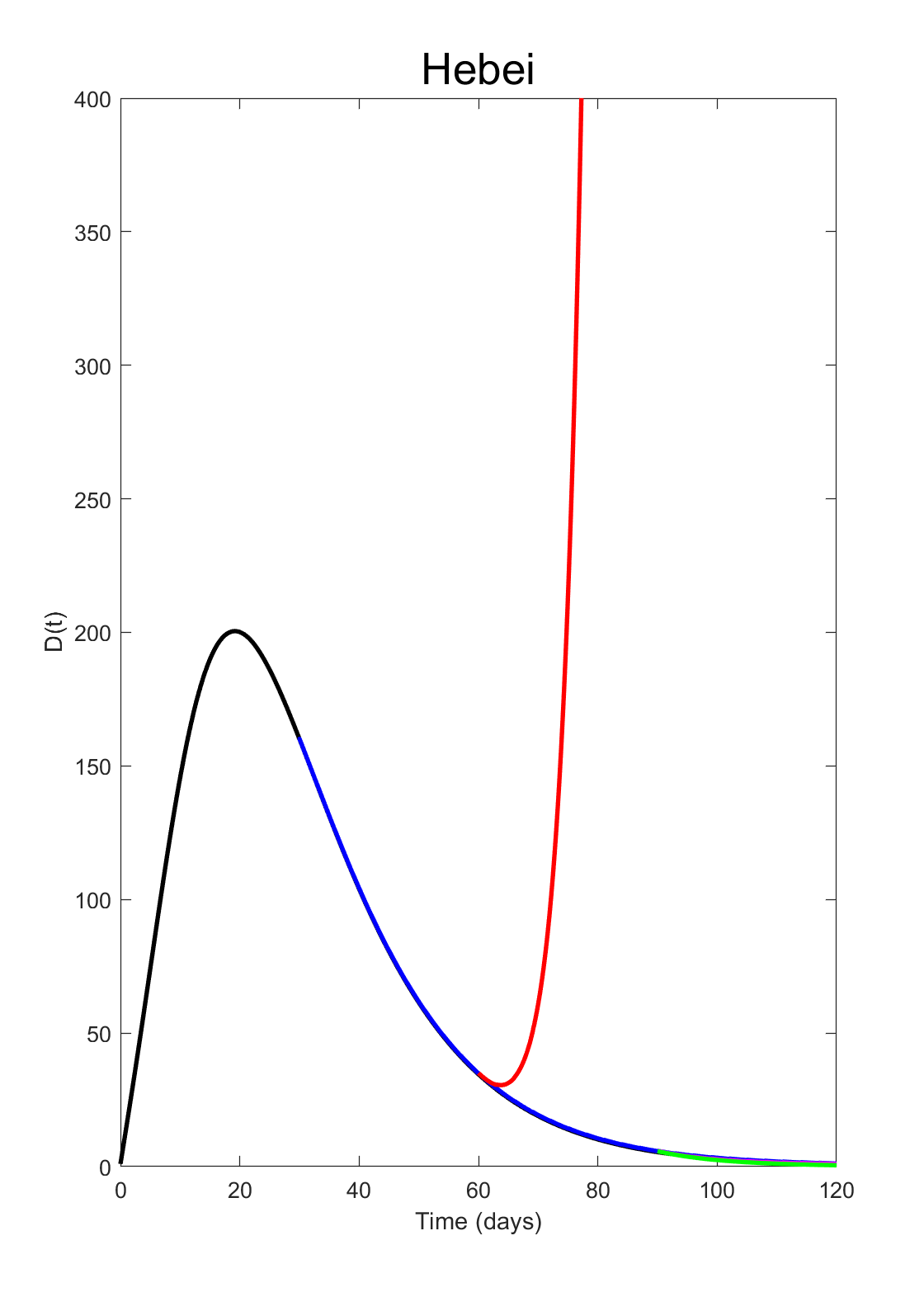}
\end{minipage}%
}%
\subfigure{
\begin{minipage}[t]{0.25\textwidth}
\centering
\includegraphics[width=3.6cm]{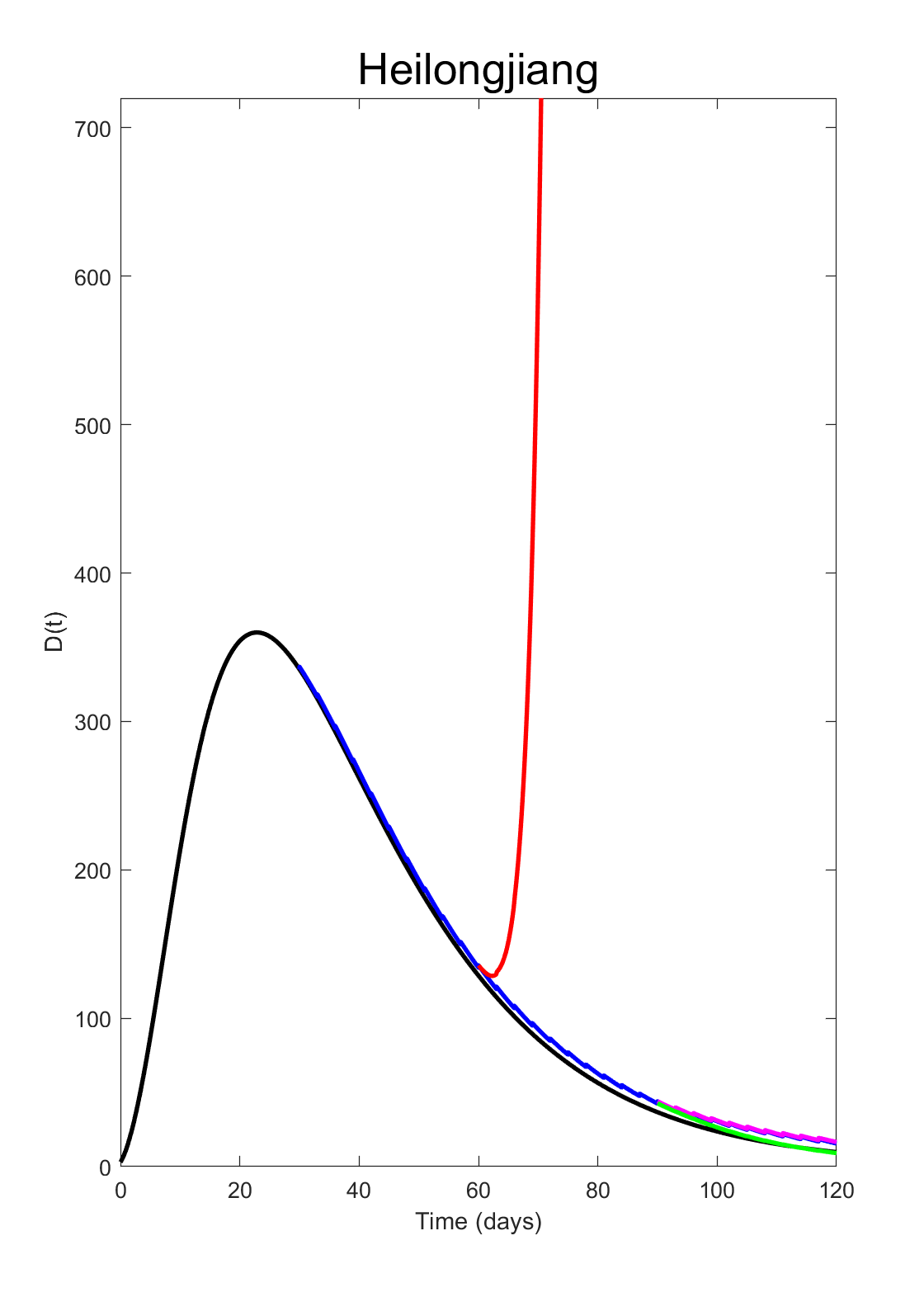}
\end{minipage}%
}%
\subfigure{
\begin{minipage}[t]{0.25\textwidth}
\centering
\includegraphics[width=3.6cm]{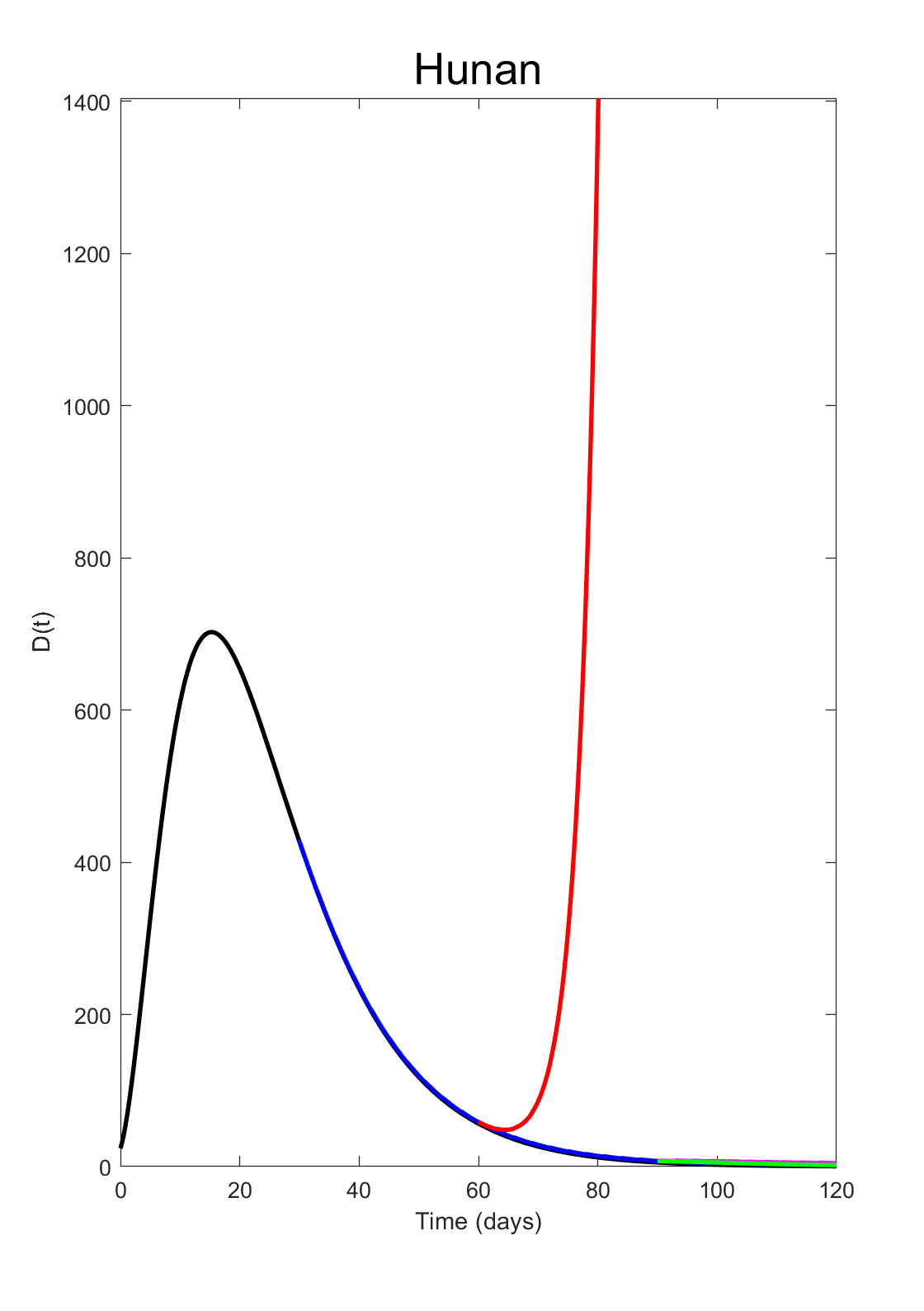}
\end{minipage}%
}%
\subfigure{
\begin{minipage}[t]{0.25\textwidth}
\centering
\includegraphics[width=3.6cm]{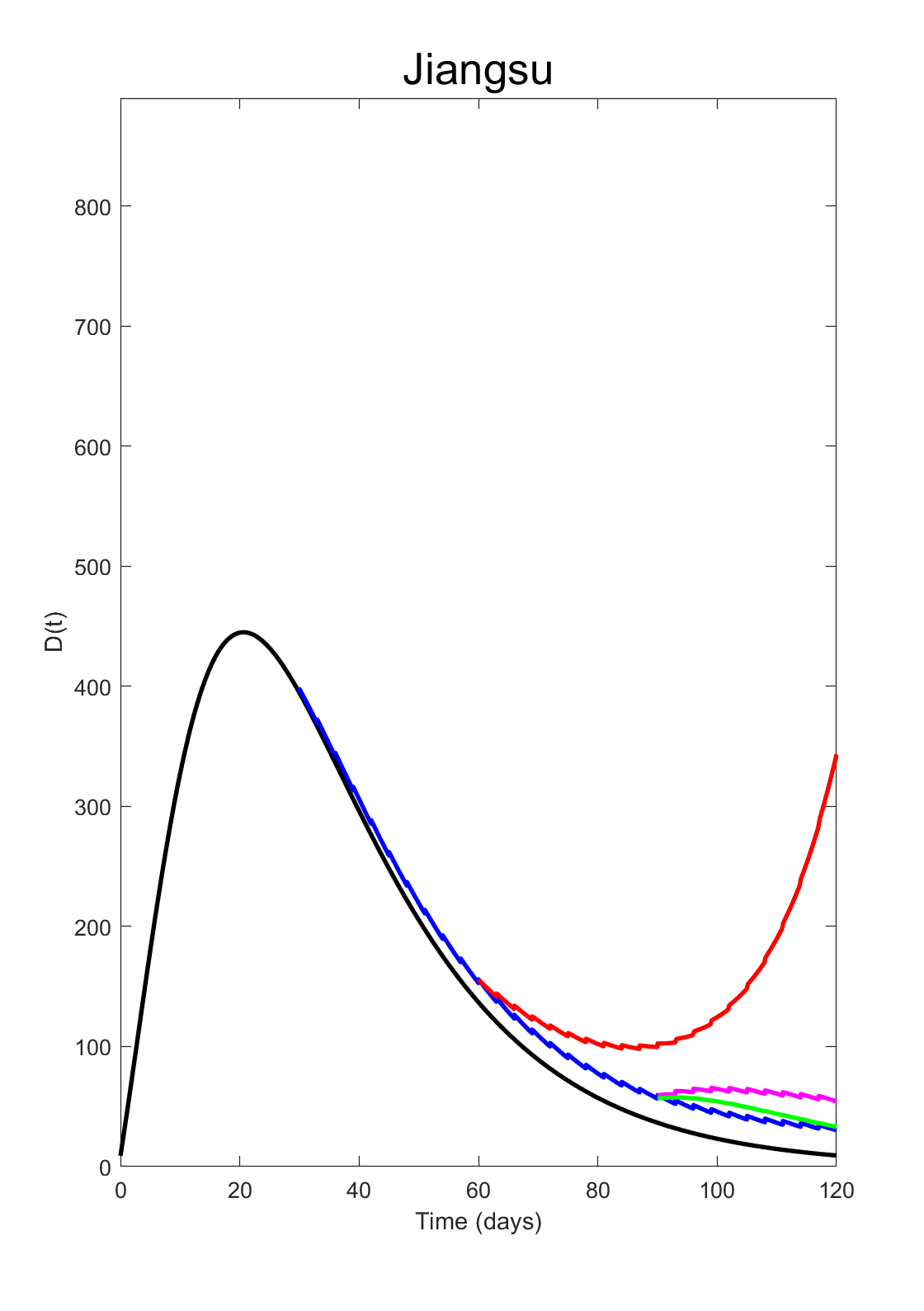}
\end{minipage}%
}%
\vskip -10pt

\subfigure{
\begin{minipage}[t]{0.25\textwidth}
\centering
\includegraphics[width=3.6cm]{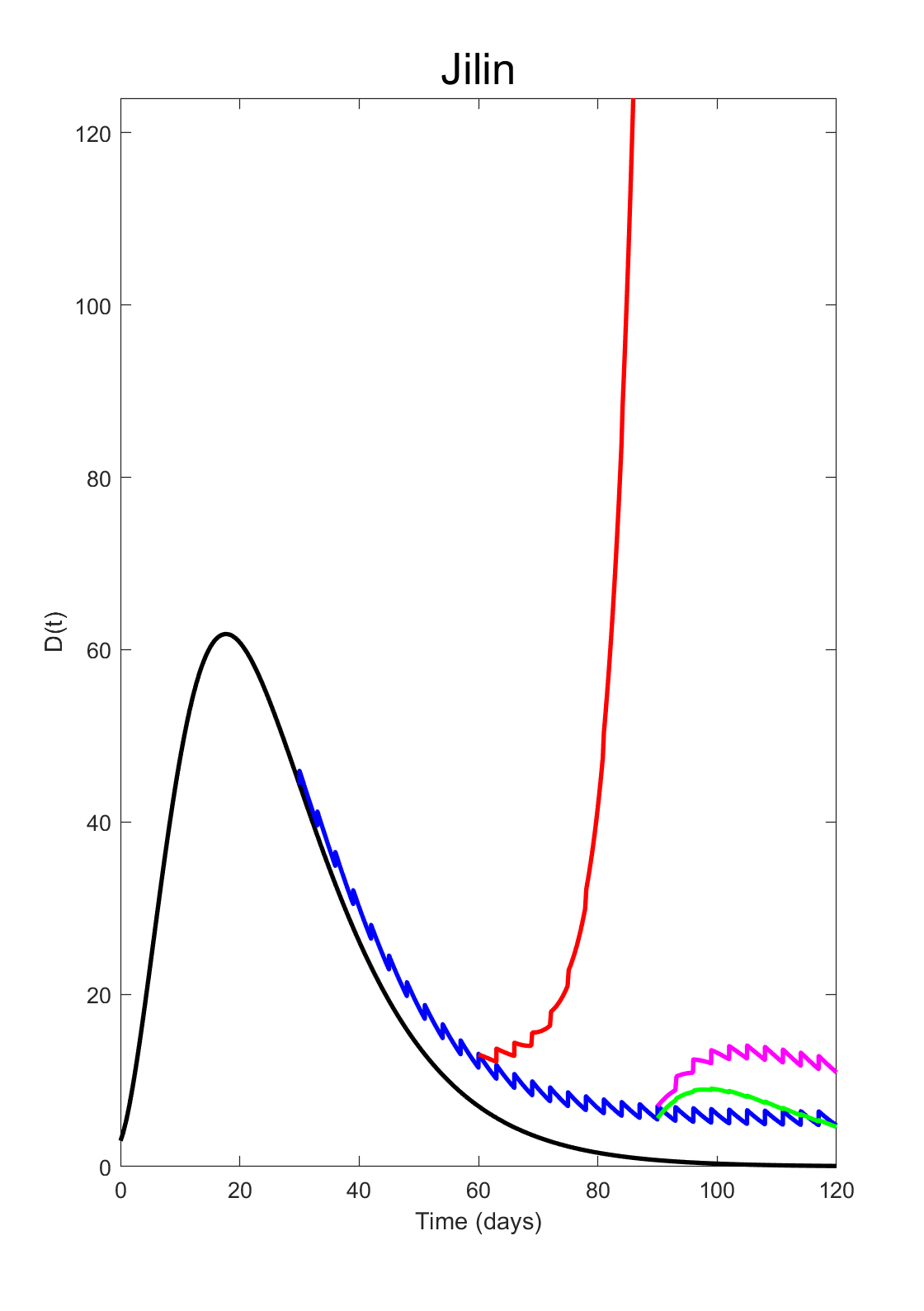}
\end{minipage}%
}%
\subfigure{
\begin{minipage}[t]{0.25\textwidth}
\centering
\includegraphics[width=3.6cm]{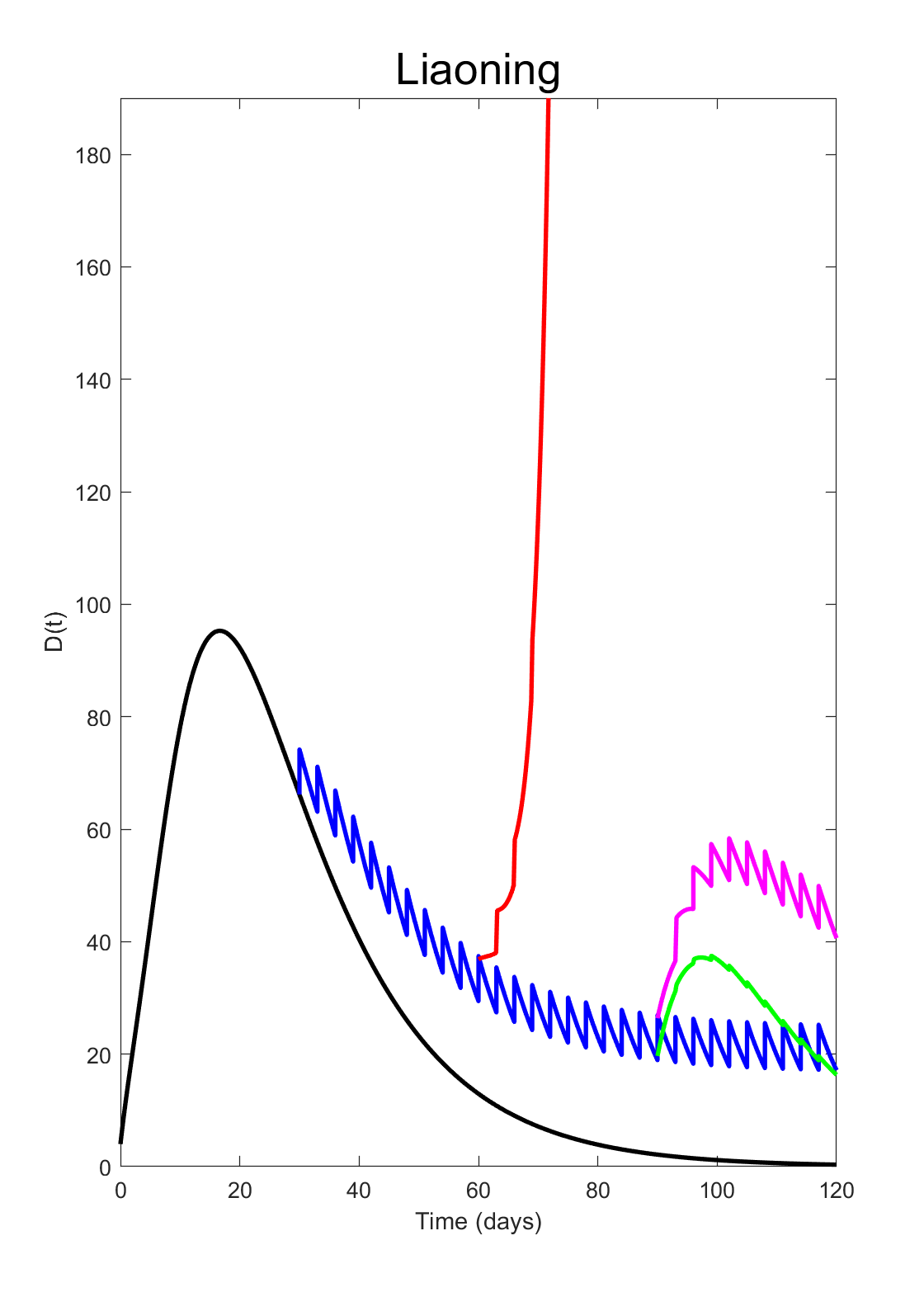}
\end{minipage}%
}%
\subfigure{
\begin{minipage}[t]{0.25\textwidth}
\centering
\includegraphics[width=3.6cm]{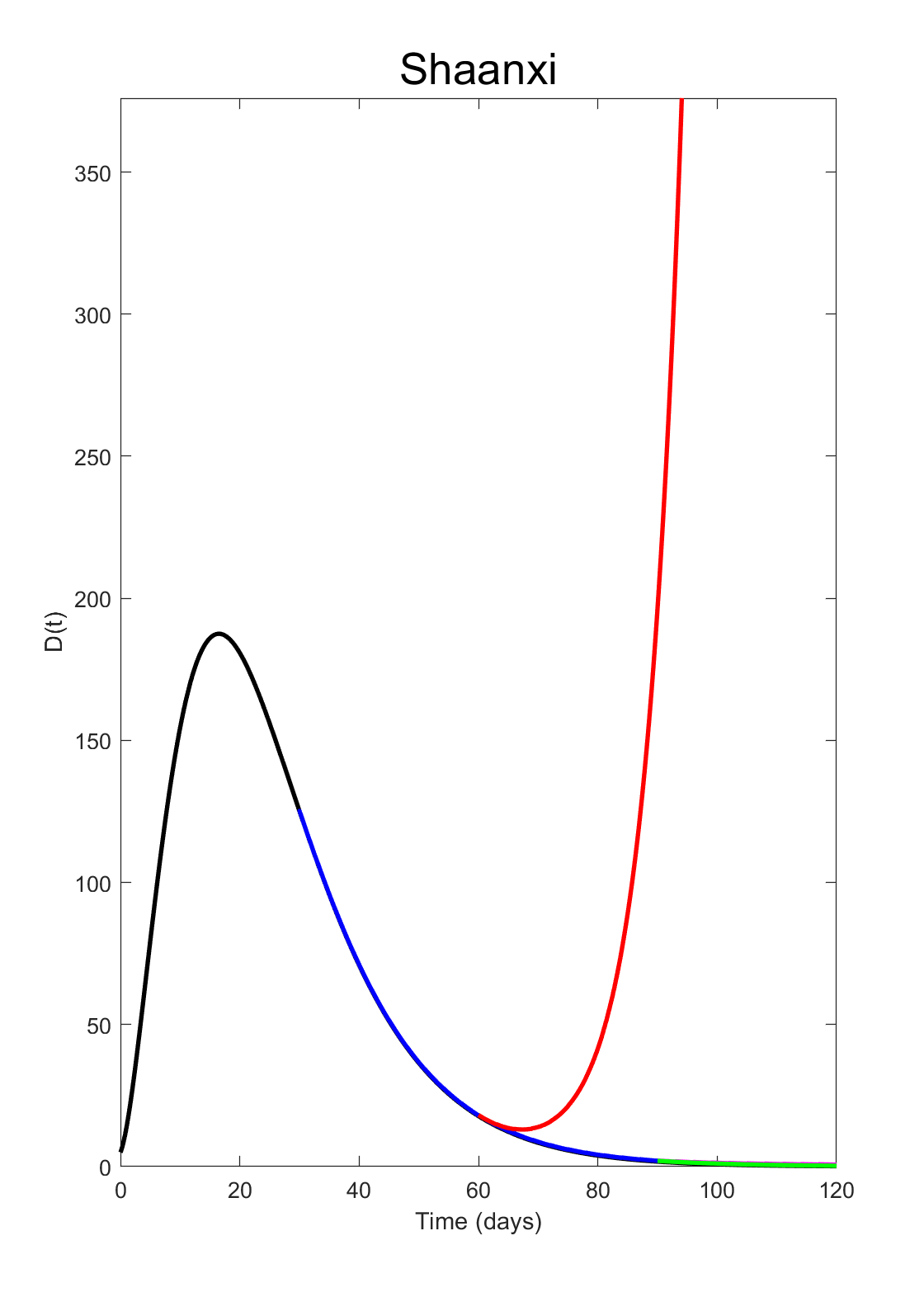}
\end{minipage}%
}%
\subfigure{
\begin{minipage}[t]{0.25\textwidth}
\centering
\includegraphics[width=3.6cm]{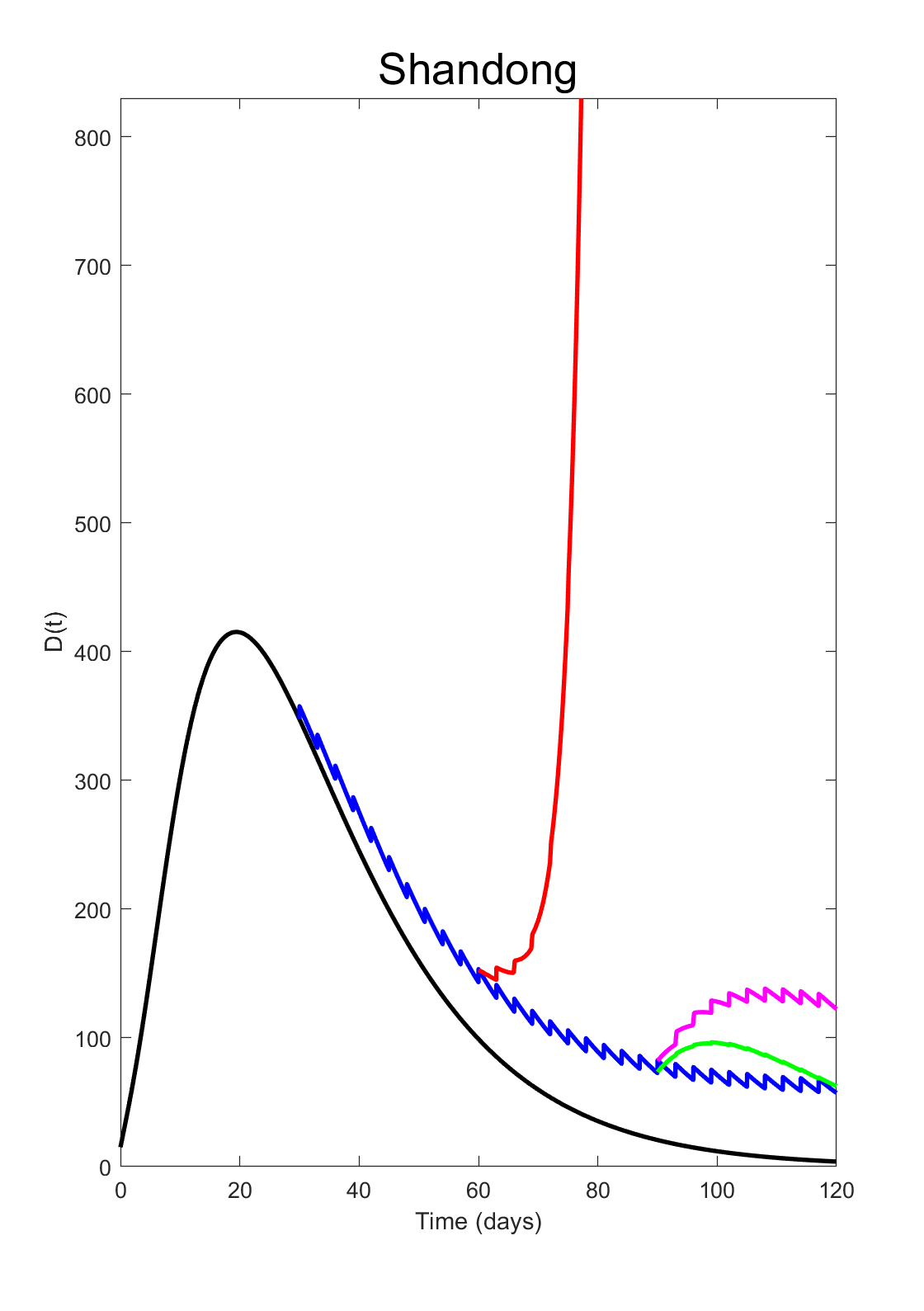}
\end{minipage}%
}%
\vskip -10pt

\subfigure{
\begin{minipage}[t]{0.25\textwidth}
\centering
\includegraphics[width=3.6cm]{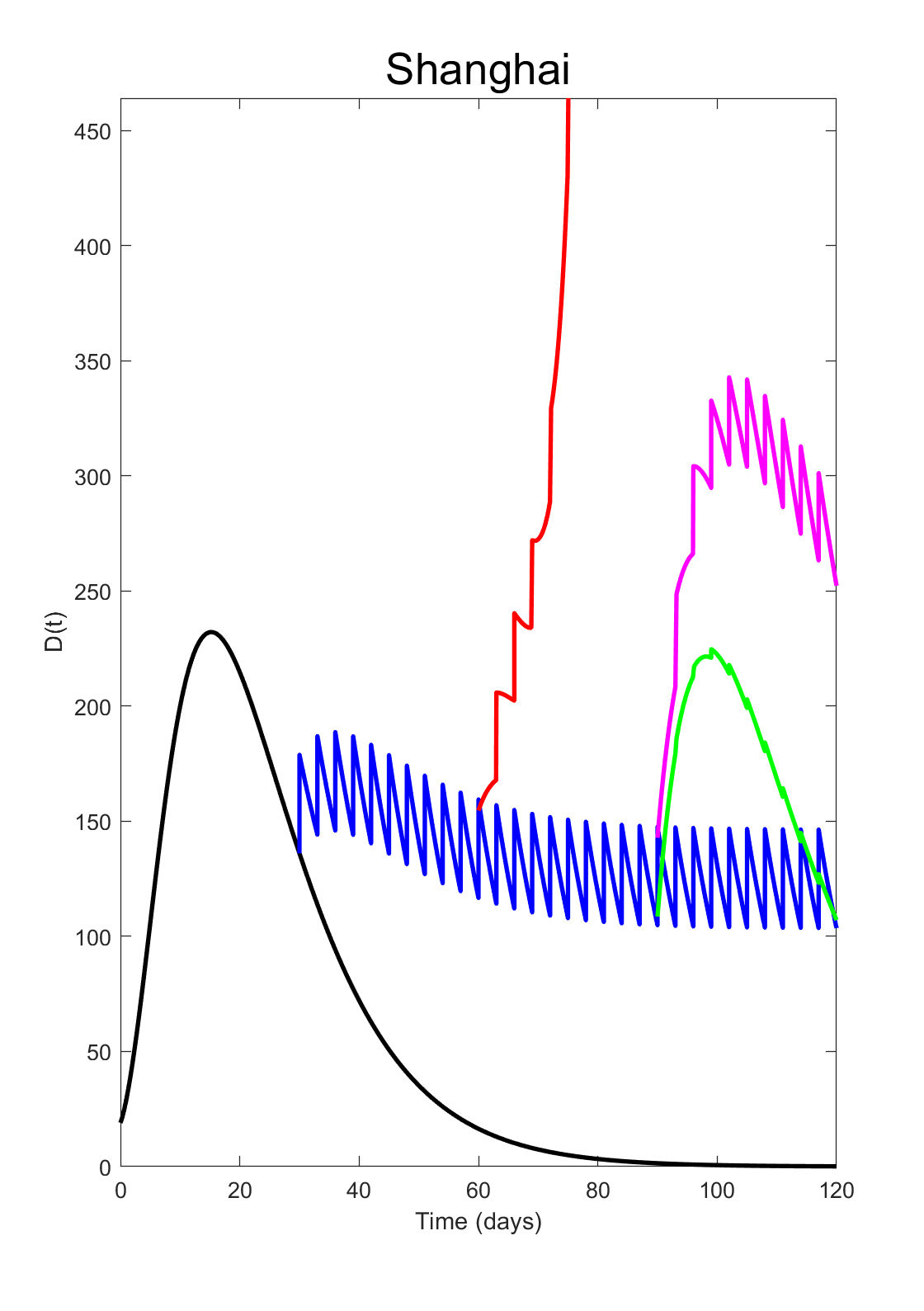}
\end{minipage}%
}%
\subfigure{
\begin{minipage}[t]{0.25\textwidth}
\centering
\includegraphics[width=3.6cm]{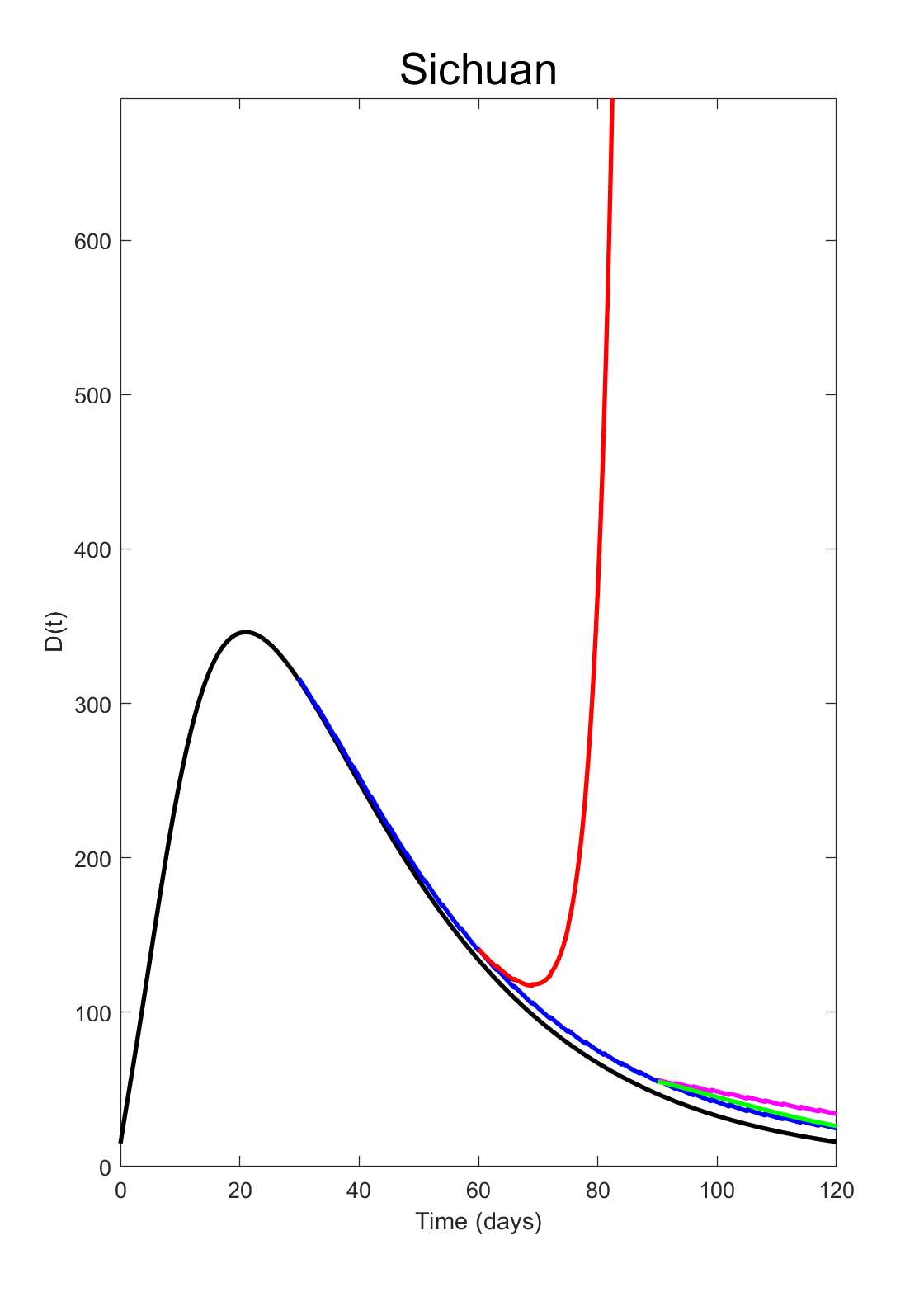}
\end{minipage}%
}%
\subfigure{
\begin{minipage}[t]{0.25\textwidth}
\centering
\includegraphics[width=3.6cm]{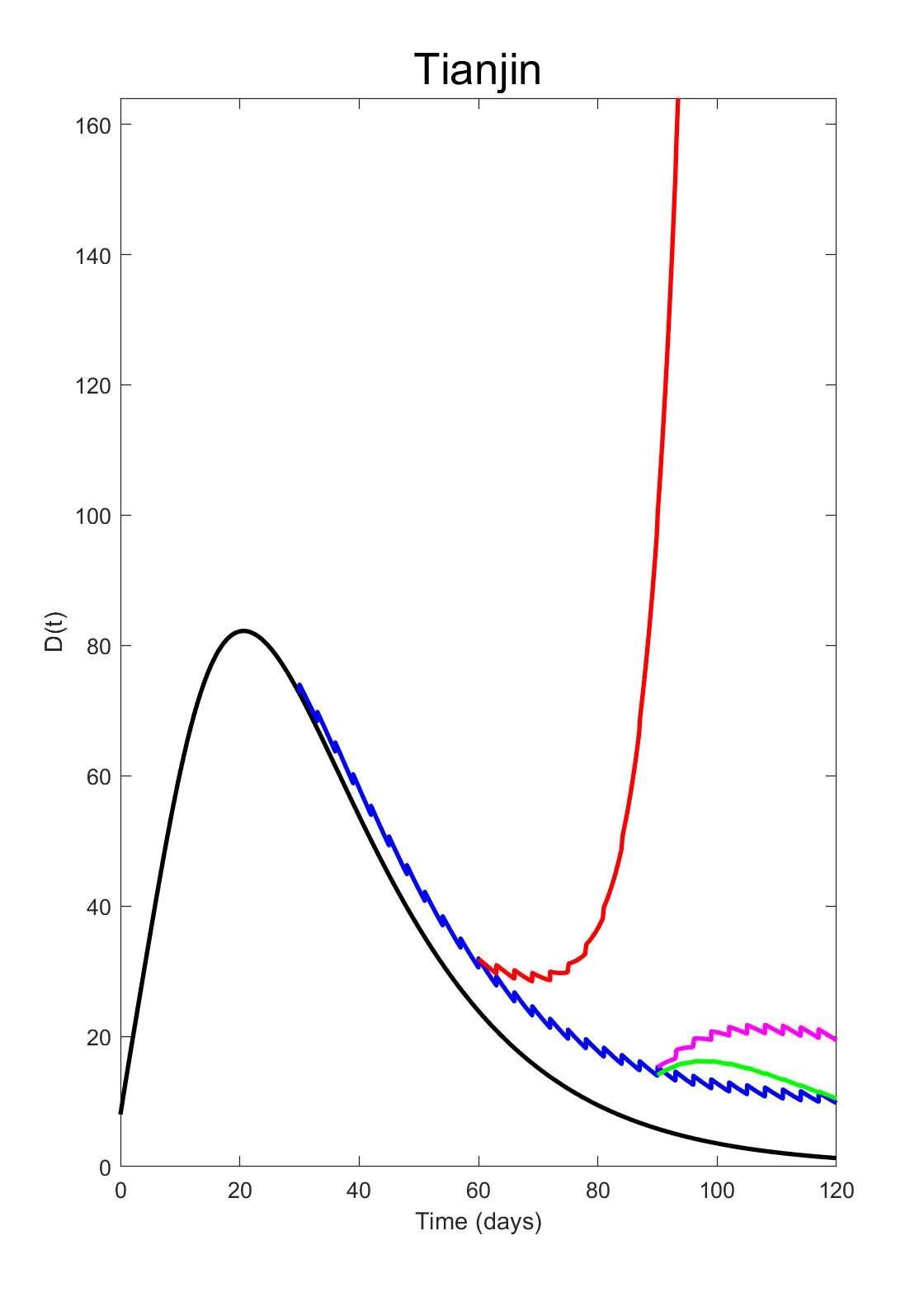}
\end{minipage}%
}%
\subfigure{
\begin{minipage}[t]{0.25\textwidth}
\centering
\includegraphics[width=3.6cm]{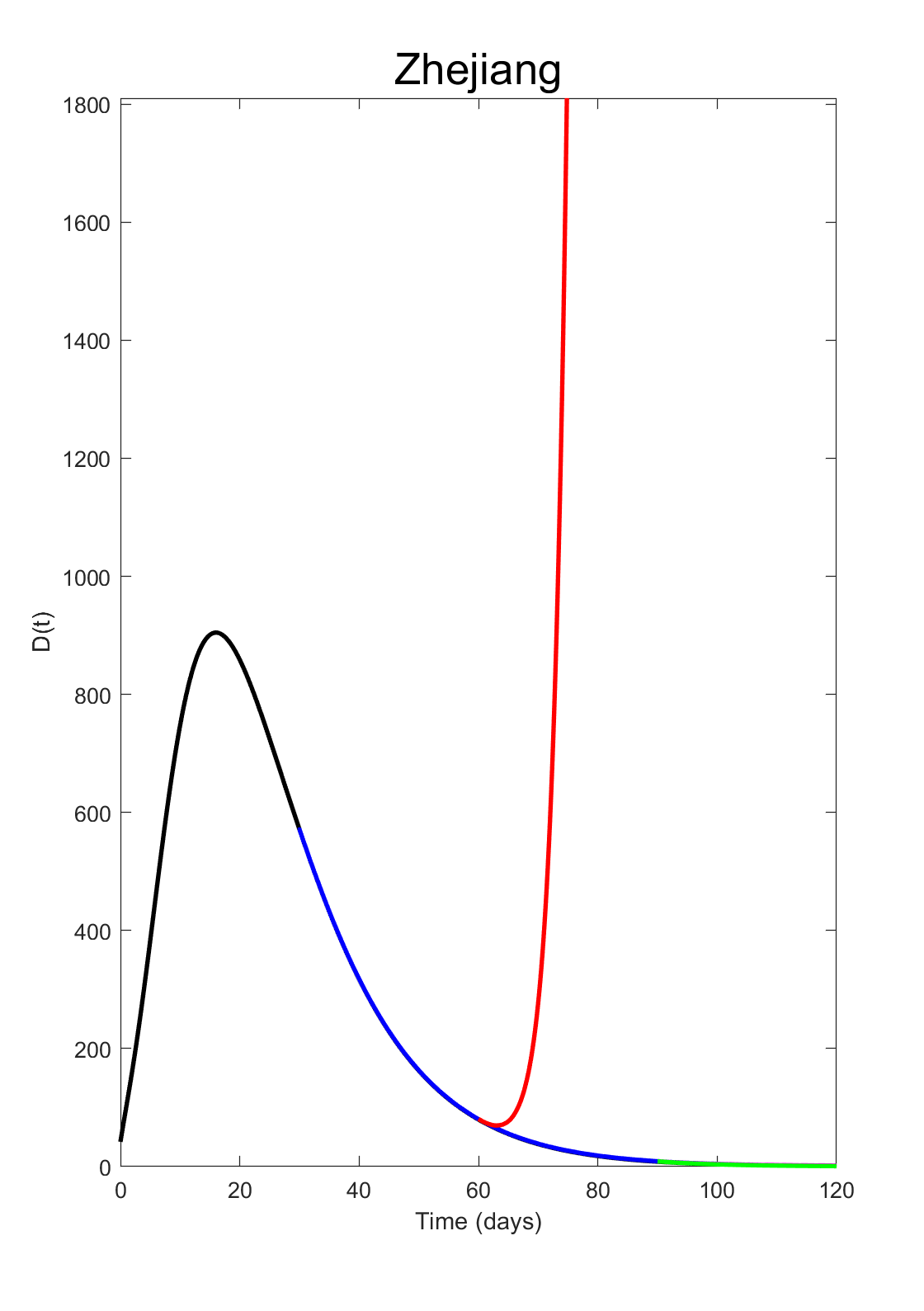}
\end{minipage}%
}%
\vskip -10pt

\subfigure{
\begin{minipage}[t]{\textwidth}
\centering
\includegraphics[width=0.8\textwidth]{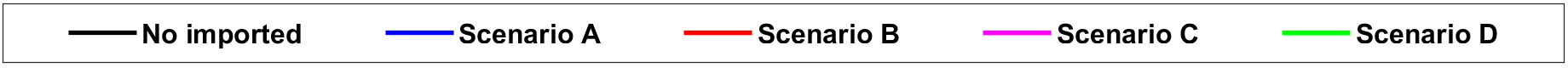}
\end{minipage}%
}%

\centering
\caption{Simulation results}\label{fig_result}
\end{figure}
China, there are still a few domestic infected cases, we assume that all the domestic
exposed and infected cases will be cleaned up before Apr 22, it means $E(t), A(t)$ and $I(t)$ will be keeping $0$ after that day.
We apply the model in \cite{EJDE} to simulate the no imported scenario and the first $30$ days of above four scenarios.
We combine Model~(\ref{model1}) for the next $90$ days to finish the simulation for Scenario A. We use Model~(\ref{model1}) for second $30$ days and
Model~(\ref{model2}) for the following $60$ days with condition $E(t)+A(t)+I(t)\neq0$ to describe Scenario B. For Scenario C, we use the combination of
$60$ days for Model~(\ref{model1}) and $30$ days for Model~(\ref{model2}) to simulate Scenario~C with condition $E(t)+A(t)+I(t)=0$.
On Mar 26, CAAC announced that each domestic airline can keep only one route to each country and each route cannot operate more than one shift per week, each foreign airline can keep only one route to China and the weekly operation shift cannot exceed one \cite{minhang}. This strategy caused a sharply drop of international flights from 1447 to 131 shifts per week. Scenario~D describes the impact of this strategy with dropping $90\%$ of impulsive imported population. We refer the parameters estimated in \cite{EJDE} for simulation (see Table~\ref{table_para}).

We summarize the results in Figure~\ref{fig_result}, we can clearly see that, if the home quarantine strategy is keeping (Scenario A), the impact of imported cases is limited, $D(t)$ will stay at a relatively low level in most selected provinces. But for high imported regions, such as Beijing, Liaoning and Shanghai, the medical burden is heavy and enduring. Notice that, if the domestic infected cases hasn't been cleared, it is very dangerous to resume to work and may cause new outbreak in each selected province. In Scenario C and D, we choose the beginning day on April 22, 2020. According to daily report, we assume that from that day, the infected population is vanished. At that time, local people return to their normal life and the imported population will be isolation for at least $14$ days. We show different imported risk simulation. Under this strategy, for most provinces, the control effect is acceptable. But in high imported risk regions, the medical resources are still strained. Especially in Shanghai, the peak value of $D(t)$ is much higher than the previous outbreak. We also calculate the increment of $AMR$ for C and D scenarios, see Table~\ref{table_AMR_IIIIV}. Since the Chinese government has pushed the tightened immigration policy, it will decrease $D(t)$ compared with normal immigration policy. And the corresponding medical burden is also decreased.

Notice that, many imported individuals need two or more transfer to get their destination. Due to the strong transmission ability of COVID-19, the transportation of imported population may cause new local infections. Considering Scenario B, if local people are infected by the imported cases and we don't keep home quarantine strategy, the potential outbreak inside China will be much horrible. To avoid this, it is suggested that, nucleic acid testing should be applied to each imported individuals and isolating the imported population immediately without any more domestic transfer.
\begin{table}[htbp]
\centering
\caption{$AMR$~estimation for Scenario~ C and D}\label{table_AMR_IIIIV}
    \resizebox{0.8\textwidth}{!}{
 \begin{tabular}{c|c|c|c|c|c|c|c|c}
\hline\hline
 & \textbf{BJ} & \textbf{CQ} & \textbf{FJ} & \textbf{GD} & \textbf{HB} & \textbf{HL} & \textbf{HN} & \textbf{JS} \\
\hline
Scenario C & 5079$r$ & 83$r$ & 1072$r$ & 1658$r$ & 74$r$ & 826$r$ & 186$r$ & 1840$r$ \\
Scenario D & 3254$r$ & 57$r$ & 698$r$ & 1067$r$ & 63$r$ & 668$r$ & 133$r$ & 1446$r$ \\
\hline
 & \textbf{JL} & \textbf{LN} & \textbf{SX} & \textbf{SD} & \textbf{SH} & \textbf{SC} & \textbf{TJ} & \textbf{ZJ} \\
\hline
Scenario C & 361$r$ & 1450$r$ & 36$r$ & 3701$r$ & 8624$r$ & 1344$r$ & 597$r$ & 119$r$ \\
Scenario D & 220$r$ & 865$r$ & 27$r$ & 2534$r$ & 5298$r$ & 1205$r$ & 428$r$ & 99$r$ \\

\hline\hline
\end{tabular}}
\end{table}

\section{Conclusions}
In this paper, we establish two epidemic models with pulse to describe imported population from other countries. Under different control strategies, the trends and medical burden of COVID-19 are given. If we keep home quarantine, the impact is limited. After people go back to normal life, strict immigration control is pre-requisite and domestic infected population must be vanished. Otherwise, the scale of the new outbreak will be even worse. Once the domestic infected population is clear, isolation for imported population only is acceptable. But some high imported risk provinces must prepare for long-term control measures. The imported COVID-19 cases will require high medical investment for a long time.

\section*{Acknowledgments}
This work was partially supported by National Natural Science Foundation of China (Grant No. 41704116, 11901234, 11926104),
Jilin Provincial Excellent Youth Talents Foundation (Grant No. 20180520093JH),
Scientific Research Project of Education Department of Jilin Province (Grant No. JJKH20200933KJ), Scientific Research Project of Shanghai Science and Technology Commission (Grant No. 19511132000).

\newpage
\section*{Appendix}
\renewcommand\arraystretch{1.2}
\begin{table}[htp]
\centering
\begin{threeparttable}
\caption{Imported population estimation~-~Part I}\label{table_import1}
\begin{tabular}{c|c|c|c|c|c|c|c|c|c}
\hline
\hline
\multicolumn{2}{c|}{\textbf{Imported population}}	&	\textbf{BJ}	&	\textbf{CQ}	&	\textbf{FJ}	&	\textbf{GD}	&	\textbf{HB}	&	\textbf{HL}	&	\textbf{HN}	&	\textbf{JS}	\\
\hline
\multicolumn{2}{c|}{$S^{i}$}	&	19914	&	516	&	4627	&	7501	&	248	&	2379	&	615	&	3750	\\
\multicolumn{2}{c|}{$E^{i}$}	&	415	&	9	&	89	&	150	&	3	&	33	&	14	&	65	\\
\multicolumn{2}{c|}{$A^{i}$}	&	14	&	1	&	4	&	6	&	1	&	2	&	1	&	3	\\
\multicolumn{2}{c|}{$I^{i}$}	&	26	&	1	&	6	&	10	&	1	&	3	&	1	&	5	\\
\hline
\hline
\textbf{Country}	&	\textbf{Airports}	&\multicolumn{8}{c}{\textbf{Number of flights per week}}\\
\hline
\multirow{4}{*}{Korea}	&	ICN	&	31	&	1	&	11	&	17	&		&	15	&		&	10	\\
	&	GMP	&	3	&		&		&		&		&		&		&		\\
	&	CJU	&		&		&		&		&		&		&		&	2	\\
	&	PUS	&	12	&		&		&		&		&		&		&		\\
\hline
\multirow{5}{*}{Japan}	&	KIX	&	10	&		&	2	&	5	&		&	1	&	3	&	8	\\
	&	HND	&	21	&		&		&	14	&		&		&		&		\\
	&	NGO	&	2	&		&		&		&		&		&		&		\\
	&	CTS	&	2	&		&		&		&		&		&		&		\\
	&	NRT	&		&		&	15	&		&		&	3	&		&	3	\\
\hline
Iran	&	IKA	&	3	&		&		&	5	&		&		&		&		\\
\hline
France	&	CDG	&	6	&		&		&	4	&		&		&		&		\\
\hline
\multirow{2}{*}{Germany}	&	FRA	&	7	&		&		&		&		&		&		&		\\
	&	MUC	&	3	&		&		&		&		&		&		&		\\
\hline
\multirow{2}{*}{Spain}	&	BCN	&	2	&		&		&		&		&		&		&		\\
	&	MAD	&	3	&		&		&		&		&		&		&		\\
\hline
\multirow{2}{*}{Italy}	&	FCO	&	7	&	2	&		&		&		&		&		&		\\
	&	LIN	&	8	&		&		&		&		&		&		&		\\
\hline\hline
\textbf{Country}	&	\textbf{Ports}	&\multicolumn{8}{c}{\textbf{Number of ships per week}}\\
\hline
Korea	&	SKINC	&		&		&		&		&	2	&		&		&	2	\\
\hline
\hline
\end{tabular}
\begin{tablenotes}
\footnotesize
\item[*]BJ: Beijing; CQ: Chongqing; FJ: Fujian; GD: Guangdong; HE: Hebei; HL: Heilongjiang; HN: Henan; JS: Jiangsu.
\end{tablenotes}
\end{threeparttable}
\end{table}

\renewcommand\arraystretch{1.2}
\begin{table}[htp]
\centering
\begin{threeparttable}
\caption{Imported population estimation~-~Part II}\label{table_import2}
\begin{tabular}{c|c|c|c|c|c|c|c|c|c}
\hline
\hline
\multicolumn{2}{c|}{\textbf{Imported population}}	&	\textbf{JL}	&	\textbf{LN}	&	\textbf{SX}	&	\textbf{SD}	&	\textbf{SH}	&	\textbf{SC}	&	\textbf{TJ}	&	\textbf{ZJ}	\\
\hline
\multicolumn{2}{c|}{$S^{i}$}	&	3012	&	11521	&	208	&	17014	&	42698	&	1132	&	2354	&	520	\\
\multicolumn{2}{c|}{$E^{i}$}	&	31	&	160	&	2	&	197	&	894	&	22	&	27	&	5	\\
\multicolumn{2}{c|}{$A^{i}$}	&	1	&	6	&	1	&	7	&	32	&	1	&	1	&	1	\\
\multicolumn{2}{c|}{$I^{i}$}	&	3	&	12	&	1	&	16	&	58	&	2	&	3	&	1	\\
\hline
\hline
\textbf{Country}	&	\textbf{Airports}	&\multicolumn{8}{c}{\textbf{Number of flights per week}}\\
\hline
\multirow{5}{*}{Korea}	&	ICN	&	27	&	32	&	2	&	80	&	20	&	3	&	2	&	5	\\
	&	GMP	&		&		&		&		&	19	&		&		&		\\
	&	CJU	&		&		&		&		&	7	&		&		&		\\
	&	PUS	&		&	2	&		&	7	&	10	&		&		&		\\
	&	TAE	&		&		&		&		&	2	&		&		&		\\
\hline
\multirow{10}{*}{Japan}	&	KIX	&		&	5	&		&	3	&	28	&		&		&		\\
	&	HND	&		&		&		&		&	35	&		&		&		\\
	&	NGO	&		&		&		&		&	14	&		&	2	&		\\
	&	CTS	&		&		&		&		&	7	&		&		&		\\
	&	NRT	&	1	&	14	&		&	4	&	38	&	2	&		&		\\
	&	FUK	&		&	2	&		&		&	5	&		&		&		\\
	&	HIJ	&		&	1	&		&		&	2	&		&		&		\\
	&	KMQ	&		&		&		&		&	1	&		&		&		\\
	&	SDJ	&		&		&		&		&	1	&		&		&		\\
	&	OKA	&		&		&		&		&	2	&		&		&		\\
\hline
Iran	&	IKA	&		&		&		&		&	5	&		&		&		\\
\hline
France	&	CDG	&		&		&		&		&	7	&		&		&		\\
\hline
\multirow{2}{*}{Germany}	&	FRA	&		&		&		&		&	7	&	2	&		&		\\
	&	MUC	&		&		&		&		&	1	&		&		&		\\
\hline
\multirow{2}{*}{Italy}	&	FCO	&		&		&		&		&	14	&		&		&		\\
	&	LIN	&		&		&		&		&	12	&		&		&		\\
\hline\hline
\textbf{Country}	&	\textbf{Ports}	&\multicolumn{8}{c}{\textbf{Number of ships per week}}\\
\hline
\multirow{3}{*}{Korea}	&	SKINC	&		&	28	&		&	28	&		&		&	14	&		\\
	&	KRPTK	&		&		&		&	14	&		&		&		&		\\
	&	SKKUN	&		&		&		&	7	&		&		&		&		\\
\hline
\multirow{3}{*}{Japan}	&	JPOSK	&		&		&		&		&	2	&		&		&		\\
	&	JPKOB	&		&		&		&		&		&		&	1	&		\\
	&	JPSHI	&		&		&		&	2	&		&		&		&		\\

\hline
\hline
\end{tabular}
\begin{tablenotes}
\footnotesize
\item[*]JL: Jilin; LN: Liaoning; SX: Shaanxi; SD: Shandong; SH: Shanghai; SC: Sichuan; TJ: Tianjin; ZJ: Zhejiang.
\end{tablenotes}
\end{threeparttable}
\end{table}

\begin{table}[htbp]
\centering
\caption{Parameters for simulation~\cite[Table~8, 9]{EJDE}}\label{table_para}
    \resizebox{\textwidth}{!}{
	\begin{tabular}{c|c|c|c|c|c|c|c|c}
\hline\hline																
	Parameter	&	Beijing	&	Chongqing	&	Fujian	&	Guangdong	&	Hebei	&	Heilongjiang	&	Henan	&	Jiangsu	\\
\hline	$\beta$	&	9.20E-08	&	3.63E-08	&	9.65E-08	&	2.71E-08	&	2.71E-08	&	7.93E-08	&	3.57E-08	&	5.00E-09	\\
	$\theta$	&	0.011	&	0.08	&	0.09	&	0.005	&	0.03	&	0.2	&	0.2	&	0.2	\\
	$p$	&	 1/4	&	1/6.4	&	 1/3	&	 1/3	&	 1/5	&	 1/3.5	&	 1/3	&	 1/5	\\
	$\lambda$	&	  1/60	&	  1/60	&	  1/60	&	  1/60	&	  1/60	&	  1/60	&	  1/60	&	  1/60	\\
	$\sigma$	&	 1/7	&	 1/7	&	 1/7	&	 1/7	&	 1/7	&	 1/7	&	 1/7	&	 1/7	\\
	$\rho$	&	0.94	&	0.99	&	0.93	&	0.91	&	0.9	&	0.95	&	0.89	&	0.98	\\
	$\epsilon_{A}$	&	  1/8 	&	  1/9 	&	  1/5 	&	  1/10	&	 1/7	&	  1/7 	&	 1/4	&	  1/10	\\
	$\epsilon_{I}$	&	 1/4	&	1/4.5	&	 1/4	&	1/3.3	&	 1/4	&	 1/5	&	 1/3	&	 1/7	\\
	$\gamma_{A}$	&	0.0996	&	0.119	&	0.0988	&	0.1075	&	0.1085	&	0.0967	&	0.0993	&	0.0474	\\
	$\gamma_{I}$	&	0.0766	&	0.0992	&	0.0882	&	0.0716	&	0.0724	&	0.0744	&	0.0764	&	0.0379	\\
	$\gamma_{D}$	&	0.0843	&	0.1487	&	0.097	&	0.1039	&	0.1049	&	0.0818	&	0.084	&	0.0455	\\
	$d_{I}$	&	0.0026	&	0.004	&	0.0022	&	0.0024	&	0.0024	&	0.0026	&	0.0026	&	0.001	\\
	$d_{D}$	&	0.0017	&	0.0033	&	0.0015	&	0.0016	&	0.0021	&	0.0018	&	0.0017	&	0.0007	\\
\hline	S(0)	&	19601400	&	27918000	&	24434200	&	96441000	&	51380800	&	22638000	&	83563500	&	70041090	\\
	Q(0)	&	1938600	&	3102000	&	14975800	&	17019000	&	24179200	&	15092000	&	12486500	&	10465910	\\
	E(0)	&	123	&	580	&	80	&	886	&	168	&	308	&	556	&	452	\\
	A(0)	&	81	&	90	&	65	&	37	&	40	&	25	&	211	&	145	\\
	I(0)	&	58	&	88	&	50	&	93	&	33	&	15	&	22	&	125	\\
	D(0)	&	26	&	27	&	10	&	51	&	1	&	3	&	9	&	9	\\
	R(0)	&	0	&	0	&	0	&	2	&	0	&	0	&	0	&	0	\\
\hline\hline																		
Parameter&	Jilin	&	Liaoning	&	Shaanxi	&	Shandong	&	Shanghai	&	Sichuan	&	Tianjin	&	Zhejiang	\\
\hline	$\beta$	&	4.78E-08	&	6.97E-08	&	2.82E-08	&	2.71E-08	&	1.26E-07	&	2.70E-08	&	1.42E-07	&	5.74E-08	\\
	$\theta$	&	0.005	&	0.01	&	0.09	&	0.2	&	0.01	&	0.11	&	0.005	&	0.1	\\
	$p$	&	 1/4	&	 1/3	&	1/4.5	&	 1/3	&	  1/3 	&	 1/3	&	 1/5	&	 1/3	\\
	$\lambda$	&	   1/60 	&	  1/60	&	  1/60	&	   1/60 	&	  1/60	&	  1/60	&	   1/60 	&	  1/60	\\
	$\sigma$	&	 1/7	&	 1/7	&	 1/7	&	 1/7	&	 1/7	&	 1/7	&	 1/7	&	 1/7	\\
	$\rho$	&	0.96	&	0.95	&	0.7	&	0.98	&	0.92	&	0.92	&	0.9	&	0.96	\\
	$\epsilon_{A}$	&	  1/6 	&	  1/10	&	  1/10	&	  1/5 	&	  1/4 	&	  1/4 	&	  1/10	&	  1/6 	\\
	$\epsilon_{I}$	&	 1/5	&	 1/5	&	 1/9	&	 1/3	&	 1/3	&	 1/3	&	 1/4	&	 1/3	\\
	$\gamma_{A}$	&	0.1029	&	0.1196	&	0.1298	&	0.0618	&	0.1213	&	0.0387	&	0.1254	&	0.119	\\
	$\gamma_{I}$	&	0.0686	&	0.0997	&	0.0999	&	0.0515	&	0.0808	&	0.0322	&	0.0836	&	0.0793	\\
	$\gamma_{D}$	&	0.096	&	0.1296	&	0.1098	&	0.0567	&	0.1132	&	0.0355	&	0.117	&	0.115	\\
	$d_{I}$	&	0.0029	&	0.0007	&	0.0009	&	0.0017	&	0.0035	&	0.0011	&	0.0036	&	0.0017	\\
	$d_{D}$	&	0.002	&	0.0007	&	0.0006	&	0.0012	&	0.0023	&	0.0007	&	0.0024	&	0.0012	\\
\hline	S(0)	&	24877352	&	30513000	&	23184000	&	90423000	&	20119200	&	75069000	&	14036400	&	52206700	\\
	Q(0)	&	2163248	&	13077000	&	15456000	&	10047000	&	4120800	&	8341000	&	1559600	&	5163300	\\
	E(0)	&	87	&	44	&	640	&	165	&	266	&	140	&	40	&	456	\\
	A(0)	&	6	&	25	&	36	&	60	&	22	&	64	&	26	&	145	\\
	I(0)	&	5	&	38	&	20	&	36	&	11	&	28	&	15	&	133	\\
	D(0)	&	3	&	4	&	5	&	15	&	19	&	15	&	8	&	42	\\
	R(0)	&	0	&	0	&	0	&	0	&	3	&	0	&	0	&	1	\\

\hline\hline																
\end{tabular}}
\end{table}


\newpage

\begin{thebibliography}{stringX}
\leftskip=-8mm
\parskip=-1mm
\small

\bibitem{chenhuijun}Chen H J, Guo J J, Wang C, Luo F, Yu X C, Zhang W, Li J F, Zhao D C, Xu D, Gong Q, Liao J, Yang H X, Hou W, Zhang Y Z. Clinical characteristics and intrauterine vertical transmission potential of COVID-19 infection in nine pregnant women: a retrospective review of medical records. {\it Lancet}, 2020, {\bf 395}(10226): 809-815.

\bibitem{chuanpiao}China ship ticketing Website. \url{http://chuan.114huoche8.com/}

\bibitem{minhang}Civil Aviation Administration of China Website. \url{http://www.caac.gov.cn/XXGK/XXGK/TZTG/}

\bibitem{gaojianjun}Gao J J, Tian Z X, Yang X. Breakthrough: Chloroquine phosphate has shown apparent efficacy in treatment of COVID-19 associated pneumonia in clinical studies. {\it BioSci. Trends}, 2020, {\bf 14}(1): 72-73.

\bibitem{ports}General Administration of Quality Supervision, Inspection and Quarantine, Standardization Administration of the People's Republic of China.
{\it Codes for China and World Main Shipping Trade Ports.} China: Standards Press, 2015.

\bibitem{guanweijie}Guan W J, Ni Z Y, Hu Y, Liang W H, Ou C Q, He J X, Liu L, Shan H, Lei C L, Hui D S C, Du B, Li L J, Zeng G, Yuen K Y, Chen R C, Tang C L, Wang T, Chen P Y, Xiang J, Li S Y, Wang J L, Liang Z J, Peng Y X, Wei L, Liu Y, Hu Y H, Peng P, Wang J M, Liu J Y, Chen Z, Li G, Zheng Z J, Qiu S Q, Luo J, Ye C J, Zhu S Y, Zhong N S. Clinical Characteristics of Coronavirus Disease 2019 in China. {\it NEJM}, 2020.

\bibitem{IATAweb}International Air Transport Association Website. \url{https://www.iata.org/}

\bibitem{nature}Jane Q. Covert coronavirus infections could be seeding new outbreaks. {\it Nature}, 2020.

\bibitem{EJDE}Jia J W, Ding J, Liu S Y, Liao G D, Li J Z, Duan B, Wang G Q, Zhang R.
Modeling the Control of COVID-19: Impact of Policy Interventions and Meteorological Factors. {\it EJDE}, 2020, {\bf 2020}(23): 1-24.

\bibitem{SCMcn}Jia J W, Liu S Y, Ding J, Liao G D, Wei Y H, Zhang R.
The impact of imported cases on the control of COVID-19 in China (in Chinese). {\it Science China: Mathematics}, 2020, preprint.

\bibitem{nhc}National Health Commission of the People's Republic of China Website. \url{http://www.nhc.gov.cn/}

\bibitem{panyang}Pan Y, Zhang D T, Yang P, Poon L L M, Wang Q Y. Viral load of SARS-CoV-2 in clinical samples. {\it Lancet Infect. Dis.}, 2020, {\bf 20}(4): 411-412.

\bibitem{prem}Prem K, Liu Y, Russell T W, Kucharski A J, Eggo R M, Davies N, Jit M, Klepac P. The effect of control strategies to reduce social mixing on outcomes of the COVID-19 epidemic in Wuhan, China: a modelling study. {\it the Lancet Public Health}, 2020.

\bibitem{shenmingwang}Shen M W, Peng Z H, Guo Y M, Xiao Y N, Zhang L. Lockdown may partially halt the spread of 2019 novel coronavirus in Hubei province, China. {\it medRxiv}, 2020.

\bibitem{tianhuaiyu}Tian H Y, Liu Y H, Li Y D, Wu C H, Chen B, Kraemer M U G, Li B Y, Cai J, Xu B, Yang Q Q, Wang B, Yang P, Cui Y J, Song Y M, Zheng P, Wang Q Y, Bjornstad O N, Yang R F, Grenfell B T, Pybus O G, Dye C. An investigation of transmission control measures during the first 50 days of the COVID-19 epidemic in China. {\it Science}, 2020, eabb6105.

\bibitem{WHOweb} World Health Organization Website.
\url{https://www.who.int/}

\bibitem{wrapp}Wrapp D, Wang N S, Corbett K S, Goldsmith J A, Hsieh C J, Abiona O, Graham B S, McLellan J S. Cryo-EM structure of the 2019-nCoV spike in the prefusion conformation. {\it Science}, 2020, {\bf 367}(6483): 1260-1263.

\bibitem{zhangsheng}Zhang S, Diao M Y, Yu W B, Pei L, Lin Z F, Chen D C. Estimation of the reproductive number of novel coronavirus (COVID-19) and the probable outbreak size on the Diamond Princess cruise ship: A data-driven analysis. {\it IJID}, 2020, {\bf 93}: 201-204.

\bibitem{Zhu2020} Zhu N, Zhang D, Wang W, {\it et al.}
A Novel Coronavirus from Patients with Pneumonia in China, 2019.
{\it NEJM}, 2020, {\bf 382}(8): 727-733.

\bibitem{zoulirong}Zou L R, Ruan F, Huang M X, Liang L J, Huang H T, Hong Z S, Yu J K, Kang M, Song Y C, Xia J Y, Guo Q F, Song T, He J F, Yen H J, Peiris M, Wu J. SARS-CoV-2 Viral Load in Upper Respiratory Specimens of Infected Patients. {\it NEJM}, 2020, {\bf 382}(12): 1177-1179.


%
%
%
%
%
%
%
%
%

\end{thebibliography}
\end{document}